\documentclass[twocolumn,twocolappendix]{aastex631}
\usepackage{amsmath}
\usepackage{amssymb}
\usepackage[utf8]{inputenc}
\usepackage[english]{babel}
\usepackage{amsfonts}
\newcommand{\gdet}{\sqrt{-g}}
\newcommand{\p}{\partial}
\usepackage{graphics}
\usepackage{graphicx}
\usepackage{epstopdf}
\usepackage{footnote}
\usepackage{txfonts}
\usepackage{lipsum}
\usepackage{float}
\usepackage{gensymb}
\usepackage{bm}

\DeclareSymbolFont{cmletters}{OML}{cmm}{m}{it}
\DeclareMathSymbol{v}{\mathalpha}{cmletters}{"76}

\definecolor{darkblue}{rgb}{0.0,0.0,0.3}
\hypersetup{colorlinks,breaklinks,
            linkcolor=darkblue,urlcolor=darkblue,
            anchorcolor=darkblue,citecolor=darkblue}

\newcommand{\rg}{r_\mathrm{g}}

\begin{document}

\title{Black Hole Collisions With Thin Accretion Disks:\\
OJ 287 and Small-Mass-Ratio Supermassive Black Hole Binary Candidates}

\author[0000-0003-0220-5723]{Sean M. Ressler}
\affiliation{Canadian Institute for Theoretical Astrophysics, University of Toronto, Toronto, On, Canada M5S 3H8}

\author[0000-0002-5427-1207]{Luciano Combi}
\affiliation{Perimeter Institute for Theoretical Physics, Waterloo, Ontario N2L 2Y5, Canada}
\affiliation{Department of Physics, University of Guelph, Guelph, Ontario N1G 2W1, Canada}

\author[0000-0002-7301-3908]{Bart Ripperda}
\affiliation{Canadian Institute for Theoretical Astrophysics, University of Toronto, Toronto, On, Canada M5S 3H8}
\affiliation{David A. Dunlap Department of Astronomy, University of Toronto, 50 St. George Street, Toronto, ON M5S 3H4}
\affiliation{Department of Physics, University of Toronto, 60 St. George Street, Toronto, ON M5S 1A7}
\affiliation{Perimeter Institute for Theoretical Physics, Waterloo, Ontario N2L 2Y5, Canada}

\author[0000-0003-0750-3543]{Xinyu Li}
\affiliation{Department of Astronomy, Tsinghua University, Beijing, 100084, China}


\begin{abstract}
OJ 287 is the best-known supermassive black hole binary candidate in the nanohertz gravitational wave band.
It exhibits periodic flares every $\sim$ 12 years, likely caused by collisions of a smaller-mass secondary with the accretion disk surrounding a larger-mass primary.  
It is therefore an important benchmark for understanding black hole binary accretion in the approaching era of space-based gravitational wave detectors and large electromagnetic surveys. 
Because the electromagnetic emission of the system is determined by a complex interplay of plasma, accretion, and radiation physics in strong gravity, numerical simulations are required for realistic modeling.
    We present the first global, three-dimensional, general relativistic magnetohydrodynamic (GRMHD) simulations of OJ 287-like systems; namely, smaller-mass secondaries colliding with a radiatively-cooled (thin) disk surrounding a larger-mass primary.
    We focus on disks with scale heights that are 10\% of the distance from the primary and binary mass ratios of $q = 0.1,0.05$, and $0.025$ using an optically-thin cooling prescription.
    We confirm the basic paradigm that impacts of the secondary on the disk can generate enough power to outshine the quiescent emission.
    The secondary also causes spiral shocks to form in the disk, enhanced accretion events, overall heating of the flow, and stochastic tilting of the disk, though these effects are small for $q<0.05$.
    Our results can be extrapolated to the parameters of OJ 287 and similar systems, an important step on the path toward fully realistic simulations of accretion onto small-mass-ratio black hole binaries and predicting electromagnetic counterparts to low-frequency gravitational wave detections.
\end{abstract}

\section{Introduction}
The blazar OJ 287 is one of the most long-studied extragalactic astronomical sources, with data dating back to the late 1800's (\citealt{Valtonen2023}, see also \citealt{Dey2019} for a review).  
In particular, it has consistently displayed optical flares every $\sim$ 12 years, which have widely been interpreted in the context of a supermassive black hole binary (SMBH) model \citep{Sillanpaa1988}.
In this model (e.g., \citealt{Valtonen2012}), an SMBH with mass $\sim$ $1.4 \times 10^{8} M_\odot$ orbits an ultramassive black hole with mass $\sim$ $1.8 \times 10^{10} M_\odot$  (where $M_\odot$ is a solar mass) and repeatedly collides with its accretion disk.
The orbit is approximately perpendicular to the plane of the disk and eccentric, with a binary separation distance that varies between $\sim$ 10--50 Schwarzschild radii of the primary black hole \citep{Dey2018}.  
As a result of the collisions, the supersonic motion of the secondary can shock-heat a fraction of the cold disk gas, producing hot, optically thick bubbles on both sides of the disk that subsequently expand and emit radiation as they cool \citep{Lehto1996,Ivanov1998}.
In order to fit the observed flaring times in OJ 287 the model requires that the secondary have an eccentric orbit that significantly precesses in the orbital plane due to general relativistic effects. 
The model has been refined over the past few decades (e.g., \citealt{Valtonen2007,Dey2018,Valtonen2023b,Zwick2023}) and has successfully predicted the timing of several flaring events (e.g., \citealt{Sillanpaa1996,Sillanpaa1996b,Valtonen2006,Valtonen2008,Valtonen2016,Laine2020}).

In addition to the optical flares that occur every $\sim$ 12 years, OJ 287 also displays more complicated short-term optical variability \citep{Pihajoki2013,Valtonen2017} as well as UV/X-ray flares not directly associated with expected disk impacts \citep{Komossa2020,MOMO4}.
These could potentially be caused by the aftermath of secondary collisions leading to enhanced accretion events, large-amplitude perturbations to the primary accretion disk, accretion onto the secondary \citep{Pihajoki2013b,Titarchuk2023,Valtonen2024b}, or magnetic reconnection events induced by the secondary \citep{Boula2025}, although these interpretations remain uncertain.
Moreover, OJ 287's polarized, relativistic jet is known to wobble significantly on $\sim$ year timescales \citep{Agudo2012}, which can also be interpreted in the binary model \citep{Valtonen2013,Dey2021}.
The jet is also likely the source of the persistent $\gamma$-ray emission with occasional flares \citep{Hodgson2017,MOMO5,Pasumarti2024,Harutyunyan2025} and the single very high energy (VHE) flare that has been observed \citep{OBrien2017,Acharyya2024} after several non-detections (e.g., \citealt{Seta2009,Archambault2016}).
Past and ongoing multi-wavelength efforts, such as the MOMO (Multiwavelength Observations and Modeling of OJ 287) campaign, have played a central role in characterizing this variability and continue to provide critical monitoring for upcoming activity (\citealt{Momo2021,MOMO5,MOMO6}; see also \citealt{Prince2021,Jormanainen2025,Harutyunyan2025}).


Recently, \citet{Komossa2023} argued that the standard precessing binary model needs to be significantly revised and that the central supermassive black hole is more likely $\sim 10^8 M_\odot$, a factor of over a hundred smaller than previous estimates, while the secondary need not be on a highly precessing orbit \citep{MOMO6}.
This argument is based on both an undetected (but purportedly predicted) flare in October 2022 and overall luminosity estimates.
Both of these points, however, have been contested on the grounds that the ``missing'' flare should actually have occurred in July of 2022 when OJ 287 was unobservable, and that the claimed luminosity discrepancy is dependent on a nonstandard calculation that would also raise issues for other active galactic nuclei (AGN, \citealt{Valtonen2023,Valtonen2023b,Valtonen2024}).  
Future study, both observational and theoretical, will be required to fully resolve this issue.
In particular, while the orbit of the secondary in the precessing binary model is exceptionally well-constrained, the structure and dynamics of the accretion disk are far less so, providing the biggest uncertainty in the model (see, e.g., \citealt{Valtonen2024}).
This question is one that is well-suited for investigation through targeted numerical simulations.

In fact, given the wealth of tightly constraining, time variable observations, OJ 287 presents a unique opportunity for global, three-dimensional simulations to study, e.g., the nonlinear evolution of black hole-disk collisions, the precise physical origin of the observed flares, the long-term effects of the secondary on the primary disk and/or jet, accretion onto the secondary and its associated emission, binary jets and their interactions, the role of magnetic reconnection (e.g., \citealt{Boula2025}), the relationship between the disk impact flares and the jet emission responsible for the blazar, as well as AGN disk dynamics in general. 
Constructing such simulations, however, poses a significant computational challenge.
In particular, they need to include general relativity with an evolving spacetime for the binary black holes, a realistic treatment of radiation and radiation transport to accurately model the observational properties of the flares, magnetohydrodynamics (MHD) to reasonably model the accretion disk around the primary, and high resolution to fully resolve the disk (since the observed luminosity implies it is radiatively efficient and likely very thin) as well as the flow near the secondary.
Because of this challenge, previous numerical work has primarily focused on local simulations that study a single collision of a black hole with a slab of gas (e.g., \citealt{Ivanov1998,Lam2025}) or on a simplified model of the disk as a collection of particles \citep{Sundelius1997,Valtonen2007,Pihajoki2013b}.
Recently, there has been a handful of global simulations of black-hole disk collisions in the radiatively inefficient regime \citep{Sukova2021,Pasham2024,Ressler2024}, but these are more relevant for low luminosity AGN with thicker disks (i.e., not OJ 287).

In this study, we build on our previous work modeling black hole binary accretion flows \citep{Combi2021,Combi2022,Ressler2024,Ressler2025} and take a further step toward realistic, global simulations of OJ 287.
We do this using a fully general relativistic (GR) treatment of magnetohydrodynamics around in-spiraling, small-mass-ratio\footnote{Here by ``small-mass-ratio'' we refer to the loosely defined range between extreme mass ratio ($\lesssim 10^{-4}$) and equal mass ratio.} binary black holes (see \citealt{Combi2021,Combi2022,Ressler2024,CombiRessler2024}) that includes radiative cooling.
Instead of a full treatment of radiation (e.g., \citealt{White2023}), we use an optically thin cooling function that drives the gas to a desired scale height (e.g., \citealt{Noble2009}).  
While this limits our ability to robustly compare to observations, it does allow us to further our understanding of the dynamical/electromagnetic properties of the system and guide future work with a more realistic radiation treatment.
Furthermore, the simulations presented here represent the first global treatment of OJ 287-like systems (that is, small-mass-ratio black hole binaries in the radiatively-efficient regime).
Our results thus have implications not just for OJ 287 but likely for other SMBH binary candidates in galactic centers. 



Small-mass-ratio SMBH binaries and/or intermediate mass black hole-SMBH binaries are of growing interest as the era of low frequency gravitational wave astronomy approaches, led by The Laser Interferometer Space Antenna (LISA, \citealt{Flanagan1998,Amaro2007,Berry2013}) and pulsar timing arrays (PTAs, \citealt{NANOGrav:2023pdq,EPTA2023,Reardon2023}).
Isolating unique electromagnetic counterparts to these systems will enable multimessenger detection and study, which could significantly advance our understanding of how SMBHs and galaxies grow and merge over cosmic time. 
In the standard precessing binary model, OJ 287 would be a prime target for PTAs \citep{Chen2018pta,Valtonen2021}, while in the alternative model proposed by \citet{Komossa2023} it would be a prime target for LISA \citep{Komossa2024}.
In either case, OJ 287 is an especially important testing ground for our analytic and numerical models of SMBH binary accretion and emission given its long-studied history.

 The flaring model of a small object colliding with an accretion disk has also been invoked to explain the newly-discovered phenomenon of quasi-periodic eruptions (QPEs) in galactic centers. 
 QPEs are large-amplitude X-ray flares with periods of hours to weeks \citep{miniutti2019nine,Giustini2020,Arcodia2021,Arcodia2024,Hernandez-Garcia2025} that have a limited duration and are often observed after a tidal disruption event of a star by the central SMBH.
While it is tempting to link these events to an orbiting secondary supermassive or intermediate mass black hole, in order to reproduce the observations the secondaries would be required to have masses $\gtrsim 10^3-10^4 M_\odot$ \citep{Linial2023} if the impact is perpendicular to the disk.
This would have problematic implications for the growth of SMBHs in light of the estimated QPE rate \citep{Arcodia2024b,Chakraborty2025,Mummery2025}. 
Instead, QPEs are more likely caused by the extreme mass ratio inspiral (EMRI) of stellar-sized objects, either stars or small black holes that collide with the disk extremely close to the plane \citep{Dai2010,xian2021x,Sukova2021,Lu2023,Franchini2023,Linial2023,Tagawa2023,Yao2025}.
That said, there is a reasonable prospect that ongoing wide-field time-domain surveys (e.g., the Legacy Survey of Space and Time by the Vera C.\ Rubin observatory, \citealt{LSST}) will reveal new quasi-periodic or otherwise transient phenomena that may be associated with accreting SMBH binaries.
Simulations like those presented here will be key in both predicting and interpreting such observations.

This letter is organized as follows.   \S \ref{sec:methods} details our numerical methods and parameter choices, \S \ref{sec:analytics} provides an overview of some analytical considerations, \S \ref{sec:results} presents our results,  \S \ref{sec:limitations} discusses how to interpret these results in the context of OJ 287-like systems and the path towards more realistic simulations, and \S \ref{sec:conc} concludes. 

We use units such that $G=c=k_{\rm B} =M= 1$ except for instructive purposes, where $G$ is the gravitational constant, $c$ is the speed of light, $k_{\rm B}$ is the Boltzmann constant, and $M$ is the mass of the primary black hole.
We measure distances in $\rg=GM/c^2=M=1$ and time in $GM/c^3=M=1$, i.e., always with respect to the primary black hole.
The secondary mass $M_{\rm s}$ is then related to the binary mass ratio $q\equiv M_{\rm s}/M$ as $M_{\rm s} = q$.
For electrodynamic quantities, we use Lorentz-Heaviside units so that a factor of $1/\sqrt{4{\rm \pi}}$ is absorbed in the definition of the magnetic field, and the square of the magnetic field strength in the co-moving frame, $b^2$, is equal to twice the magnetic pressure, $p_{\rm mag}=b^2/2$.

\section{Methods}
\label{sec:methods}
We use the GRMHD framework in {\tt Athena++} \citep{White2016,Athenapp} for all of our simulations, modified from the public version to include time dependent spacetime metrics as described in detail in \citet{Ressler2024}.
We use the approximate, superposed metric described in \citet{Combi2021} and \citet{CombiRessler2024}, which makes use of the 4th order Post-Newtonian solver {\tt CBWave} \citep{cbwaves} for the orbital trajectory of the secondary black hole.  
For simplicity, we neglect the orbital motion of the primary black hole and fix its location to the origin of the simulation; this is valid for small mass ratios.  
We solve the standard equations of GRMHD with an additional cooling term, namely
\begin{equation}
    \begin{aligned}
    &\nabla_\mu \left(\rho u^\mu\right) &= &0 \\
    &\nabla_\mu \left(T^\mu _\nu\right) &= &\Lambda_{\rm cool}u_\nu \\
    &\nabla_\mu \left(F^{*\mu \nu}\right) &= &0,
    \end{aligned}
\end{equation}
where $\rho$ is the rest mass density, $u^\mu$ is the four velocity, $T^\mu_\nu$ is the stress-energy tensor, $F^{*\mu \nu}$ is the dual of the Faraday tensor, and $\Lambda_{\rm cool}$ is the cooling function.  
We follow \citet{Noble2009} by using an artificial cooling function to drive the temperature of the gas on an orbital time scale to a desired local scale height, $H$, which we choose to be 10\% of the distance from the primary black hole $r$ (see \S \ref{sec:parameter_choices}).  
Specifically, we adopt the form:
\begin{equation}
\Lambda_{\rm cool}  = \begin{cases}
  \Omega_{\rm K} u_{\rm g} \left(Y-1 + |Y-1|\right)^{1/2}, & \text{if } T_{\rm target}\ge T_{\rm target,s} \\
  \Omega_{\rm K,s} u_{\rm g} \left(Y_{\rm s}-1 + |Y_{\rm s}-1|\right)^{1/2}, & \text{if }T_{\rm target}< T_{\rm target,s}, \
\end{cases}
\label{eq:lambda_cool}
\end{equation}
 where $P$ is the thermal pressure, $\Omega_{\rm K} = 1/(r^{1.5} + \chi)$ is the Keplerian angular velocity, $\chi$ is the dimensionless spin of the primary, $Y = T/T_{\rm target}$, $T=P/\rho$, and $T_{\rm target}$ is a function of the desired $H/r$ (see Equation 2 in \citealt{Noble2010}, corrected from \citealt{Abramowicz1997} and \citealt{Krolik1999}).
Quantities with subscript 's' refer to those evaluated with respect to the secondary black hole.  So $\Omega_{\rm K,s} =  1/\{ q[(r^\prime/q)^{1.5} + \chi_{\rm s}/q]\}$, $Y_{\rm s} = T/T_{\rm target,s}$, $T_{\rm target,s}$ is $T_{\rm target}$ evaluated for the secondary, $\chi_{\rm s}$ is the dimensionless spin of the secondary, and $r^\prime$ is the distance from the secondary in its own frame in units of the mass of the primary.
The switch on the cooling function in Equation \ref{eq:lambda_cool} for $T_{\rm target}<T_{\rm target,s}$ allows material around the secondary to heat up beyond the temperature set by the primary disk scale height when it reaches distances of $r^\prime < q r$. 
It also shortens the cooling time by a factor of $q$.
This form of $\Lambda_{\rm cool}$ effectively assumes that the Eddington ratios for both black holes are approximately the same, at least to the point that the gas around each would have the same $H/r$ ratio\footnote{In reality, we expect the accretion rate onto the secondary to be highly variable, with potentially a higher Eddington ratio than the primary during active accretion phases (e.g., \citealt{Ressler2024}).  Because our results are relatively insensitive to the particular form of the cooling function as discussed later, we neglect these complications for simplicity.}. 
When the gas is within the innermost stable circular orbit (ISCO) radius for either black hole, we evaluate $\Omega$ and $T_{\rm target}$ (or $\Omega_{\rm s}$ or $T_{\rm target,s}$) at the ISCO radius for the appropriate black hole.  
Furthermore, we set $\Lambda_{\rm cool}$ to 0 when the gas is unbound (i.e., the relativistic Bernoulli parameter is positive).
We use an ideal equation of state with adiabatic index $\gamma=5/3$.

\begin{figure}
\includegraphics[width=0.47\textwidth]{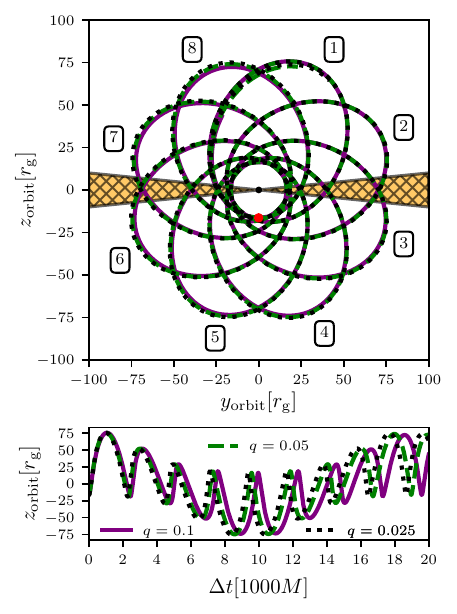}
\caption{Orbits for $q=0.1$ (solid purple), $q=0.05$ (dashed green), and $q=0.025$ (dotted black) given our chosen initial eccentricity (0.657) and binary separation (16.4 $\rg$) as determined by {\tt CBWwaves}.  
Top: Orbital trajectory of the secondary in the $y$-$z$ (orbital) plane for the first $20{, }000 M$, with the central black hole shown as a black circle and one scale height of the disk shown as the orange checkered pattern. 
The red dot denotes the initial location of the secondary, which moves clockwise, and each orbital loop is marked sequentially. 
The orbit precesses by roughly $40^\circ$ every period. 
Bottom: $z$-coordinate of the orbits as a function of time.
For different $q$ during this time period, the orbital trajectories are approximately the same, with the only significant difference being a relative time lag developing and increasing as time goes on.
}
\label{fig:orbits}
\end{figure}

We first run a thin disk simulation around an isolated black hole for $50{, }000M$ as described in detail in Appendix \ref{app:initial_condition}. 
To this disk we introduce the secondary black hole by instantaneously changing the metric (described in Appendix \ref{app:metric_change}) and running the simulations for an additional $\sim$ 10--$ 20{, }000 M$.
We use a Cartesian Kerr-Schild grid in $x$, $y$, and $z$ coordinates with nested static and adaptive mesh refinement (SMR and AMR, respectively). 
In the primary accretion disk, we use SMR to achieve roughly 30--60 cells per scale height in the $z$ direction perpendicular to the disk plane and 10--20 cells per scale height in the $x$ and $y$ directions.
The AMR is focused around the secondary black hole as described in \citet{Ressler2024} and achieves $\gtrsim 50$ cells per Bondi-Hoyle-Lyttleton radius of the secondary.\footnote{This corresponds to $\sim$ 3 cells between $r^\prime =0$ and the secondary's event horizon in the $x$ and $y$ directions (and 9 cells in the $z$ direction).
While such resolution is inadequate to study the near-horizon dynamics of the secondary accretion flow, it is more than sufficient to study its effects on the primary's disk.}

We initialize the secondary black holes with eccentricity of 0.657 and initial (and smallest) separation of 16.4$\rg$ at an inclination of 90$^\circ$ with respect to the plane of the primary accretion disk.  
This choice is motivated by the precessing binary model that has been used to explain the flares of the OJ 287 system (e.g., \citealt{Dey2019}).
Note that we place the closest approach of the  secondary's orbit slightly nearer to the primary than in the standard model so that more of the disk impacts happen where the primary disk is in inflow equilibrium ($\lesssim 25 \rg$, see Appendix \ref{app:initial_condition}).
The resulting orbits for $q=0.1, 0.05,$ and $0.025$ as computed by {\tt CBWaves} are plotted in Figure \ref{fig:orbits}. 
For all of our chosen values of $q$, the orbits precess in the orbital plane by $\sim$ 40$\degree$ per orbit. 
Note that here and throughout, we do not include cosmological redshift in the time coordinate.
For values of $q<0.1$, the mass ratio of the system mostly affects the temporal phase of the orbital precession, and this is a relatively small effect.  
The overall shape of the orbit is roughly unchanged for different $q$.

\section{Analytical Considerations}
\label{sec:analytics}
To gain insight into how our results may depend on different parameters, we first consider some simple analytical estimates.

\subsection{Energetics of Black-Hole Disk Collisions}

The total volume-integrated heating rate corresponding to the impact of the smaller secondary black hole with the primary accretion disk should roughly scale as $Q_{{\rm s}} \propto \Sigma_{\rm disk} v_{\rm orbit}^3 R_{\rm BHL}^2/H$, where $\Sigma_{\rm disk} = \int \rho r \sin(\theta) d\theta$ is the surface density of the disk, $v_{\rm orbit}$ is the orbital speed of the secondary, $R_{\rm BHL} = G M_{s}/v_{\rm orbit}^2$ is the Bondi-Hoyle-Lyttleton (BHL) radius of the secondary \citep{Hoyle1939,BHoyle1944}, and $\theta$ is the polar angle. 
This expression simplifies to $Q_{{\rm s}} \propto \Sigma_{\rm disk} q^2/(H v_{\rm orbit})$.
On the other hand, the viscous heating rate integrated over the whole accretion disk from steady-state $\alpha$-disk theory \citep{Shakura1973} is $Q_{{\rm disk}} \propto \dot M \sim [\alpha \Sigma_{\rm disk} H^2 v_{\rm orbit}/r]_{r = r_{\rm orbit}}.$
The ratio between secondary heating and disk heating is then 
\begin{equation}
\label{eq:heating_ratio}
\frac{Q_{\rm s}}{Q_{\rm disk}}  \propto \frac{1}{\alpha} \frac{1}{v_{\rm orbit}^2r_{\rm orbit}^2} \frac{q^2}{\left(H/r\right)^3}.
\end{equation}

\subsection{Relative Importance of Gravitational Torque}

The secondary can also have an effect on the disk through gravitational torque.  
We can estimate the relative importance of this effect in a simplistic model of an infinitely thin disk in the Newtonian limit by comparing the timescale for the torque to appreciably change the angular momentum of the disk, $t_{\rm torque}$, and the inflow timescale, $t_{\rm infall}$.
For $t_{\rm torque}$, we consider a parcel of gas in the disk located a distance $r$ from the primary and at azimuthal angle $\varphi$ with mass $dm = \Sigma_{\rm disk} r dr  d\varphi$ and angular momentum $dL =  dm v_{\rm kep} r$, where $v_{\rm kep}$ is the Keplerian speed.  
This parcel of gas will experience a torque of $ \mathbf{d\tau} = \mathbf{r} \times \mathbf{dF_{\rm g,s}}$, where $\mathbf{dF_{\rm g,s}} = -dm M_{\rm s} (\mathbf{r} - \mathbf{r_{\rm s}})/|\mathbf{r} - \mathbf{r_{\rm s}}|^3$, and $\mathbf{r_{\rm s}}$ is the location of the secondary.
This torque will significantly change the angular momentum of the parcel in a time $t_{\rm torque} \sim \int dL\ /\int d\tau \propto f_1(r, \mathbf{r_{\rm s}})/q$, where $f_1$ is only a function of $r$ and $\mathbf{r_{\rm s}}$ ($r$ here corresponds to the upper limit of integration for the disk) and $\mathbf{r_{\rm s}}$.
The infall time can be estimated from $\alpha$-disk theory as $t_{\rm infall} \sim r/(\alpha H^2 v_{\rm kep})$, giving
\begin{equation}
\label{eq:torque_time}
\frac{t_{\rm infall}}{t_{\rm torque}} \propto \frac{1}{\alpha}\frac{q}{\left(H/r\right)^2} f_2\left(r,\mathbf{r_{\rm s}}\right),
\end{equation}
where $f_2(r,\mathbf{r_{\rm s}})$ is another function of just $r$ and $\mathbf{r_{\rm s}}$.

It is important to note that equations \eqref{eq:heating_ratio} and \eqref{eq:torque_time} have different scalings with $q$, meaning that there is a regime where the disk can be relatively unaffected by shock heating from the secondary's impacts and yet the secondary can still exert a significant torque on the disk.  

\begin{figure*}
\includegraphics[width=0.97\textwidth]{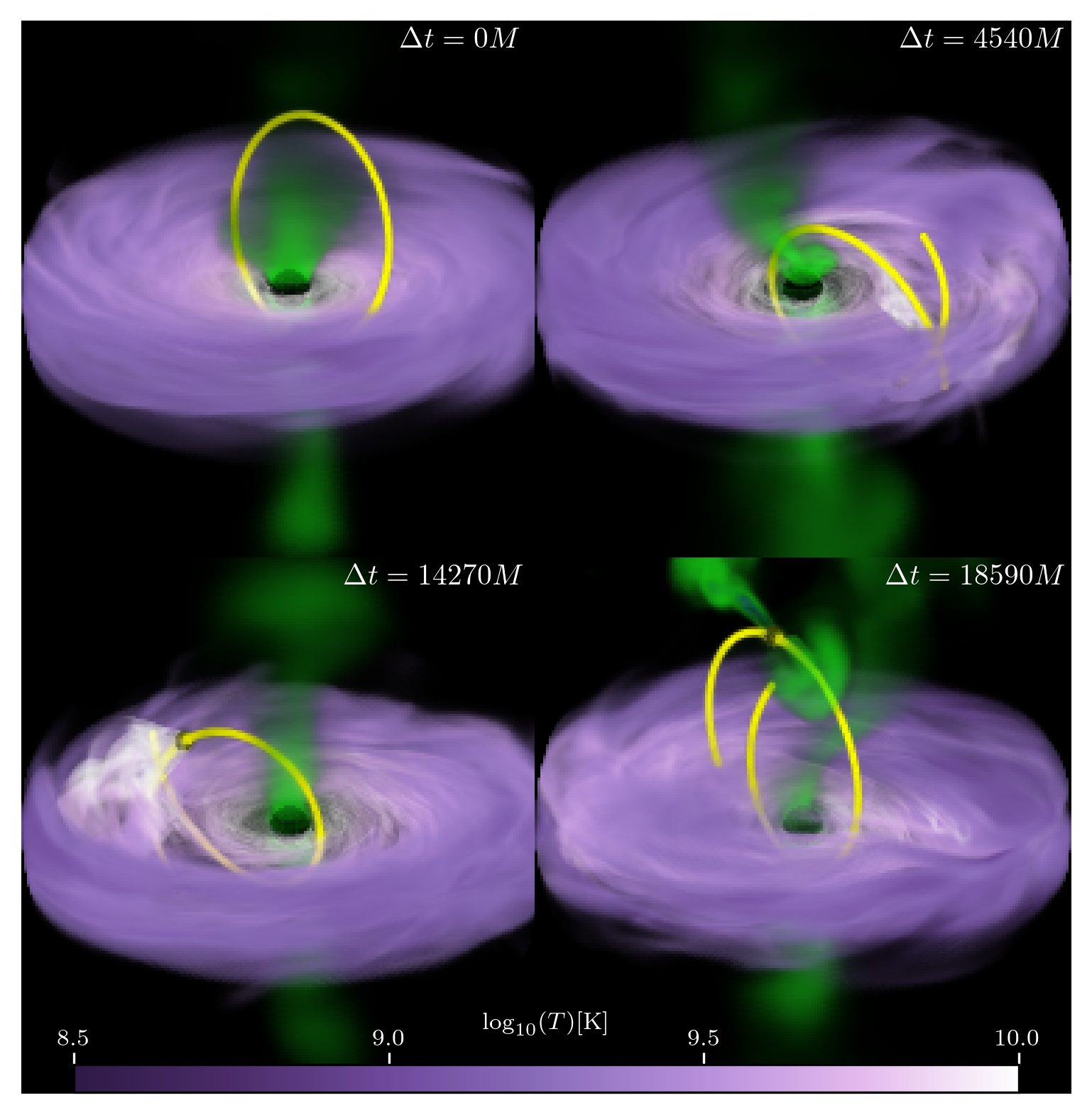}
\caption{
Volume rendering of our $q=0.1$ simulation at four different times, with the top left panel representing the time that the secondary black hole is introduced.
Disk colors show temperature (dark purple to white, logarithmic), while green marks magnetized outflow with $\sigma=1$.
The yellow lines mark the recent and future orbit for the secondary at each time (moving in the clockwise direction) and the two black spheres represent the primary and secondary black holes.
Impacts from the secondary create hot bubbles of gas on both sides of the disk while also heating up the main body of the disk and generating spiral shock waves.
Occasionally (twice in the whole simulation), a small jet is produced by the secondary (bottom right panel).
Animated version: \url{https://youtube.com/shorts/8pcuUNnYjNQ}.
}
\label{fig:3D_render}
\end{figure*}

\subsection{Parameter Choices and Extrapolating Our Results}
\label{sec:parameter_choices}
The precessing binary model for OJ 287 (e.g., \citealt{Lehto1996,Dey2019}) has a mass ratio of $q\approx0.08$ and an accretion rate onto the primary of $\sim$ 0.08 $\dot M_{\rm Edd}$ \citep{Valtonen2023}, where $\dot M_{\rm Edd}$ is the Eddington accretion rate.
The models estimate the density scale height of the disk to be $\sim 10^{15}\mathrm{\ cm}$ at the distance of secondary impact \citep{Valtonen2019}, corresponding to roughly $H/r \sim 0.01$, though there is significant uncertainty in this value \citep{Zwick2023}.
Recent radiative GRMHD simulations of luminous thin disks suggest that for a similar Eddington ratio, the gas settles into $H/r \sim 0.03$ \citep{Zhang2025}.

Generally, in order to properly resolve the MRI in thin disk simulations there needs to be $\gtrsim 30$ cells per scale height (e.g., \citealt{Noble2012}) and in order to reach inflow equilibrium even out to $\gtrsim 20 \rg$ there needs to be $\gtrsim 50{, }000M$ of integration time (e.g., \citealt{Avara2016,Liska2022,Scepi2024,Zhang2025}).
To ensure that we are always resolving the MRI while still being able to simulate several orbits of the secondary, we limit our minimum disk thickness to $H/r \sim 0.1$, which is likely at least a factor of $3$--10 thicker than reality for OJ 287, corresponding to a factor of 9--100 higher in temperature.

Based on our estimates from the previous subsection, however, we can reasonably extrapolate our results to OJ 287-like parameters.  
For example, in terms of the heating rate caused by secondary impacts, Equation \eqref{eq:heating_ratio} estimates that for our thicker $H/r=0.1$ disk, a secondary with $q \sim $ 0.05 (0.25) would have a similar relative effect on the disk as the OJ 287 secondary with $q=0.008$ acting on an $H/r =$ 0.03 (0.01) disk.
In terms of gravitational torque, Equation \eqref{eq:torque_time} implies that the pair of parameters $q\sim$ 0.09 (0.8), $H/r=0.1$ would behave similarly to the OJ 287 values of $q=0.008$, $H/r\sim$0.03 (0.01).\footnote{
Note that this extrapolation is no longer meaningful for the $q>0.2$ values where the approximation of the secondary as a perturber starts to break down.}
What this means is that for a parameter survey in $q$, even for a fixed $H/r=0.1$, we can draw meaningful conclusions for OJ 287 and other potentially similar systems.

This extrapolation, however, should be applied with caution, as realistic disks may scale differently than the simple $\alpha$-disk estimates.  
In particular, the value of $\alpha$ used in Equation \eqref{eq:heating_ratio} and \eqref{eq:torque_time} in reality can depend on the other parameters and the MHD dynamics of the flow.
Furthermore, the heating rate induced by the secondary can differ from this simple estimate.
In fact, as we will show later, we find that in our simulations $Q_{\rm s}$ scales as $\tilde \propto\ q^{1.5} $ instead of the expected $q^2$.
This adds additional uncertainty to applying our results to OJ 287.

In this work, we run simulations for $q=0.1$, $q=0.05$ and $q=0.025$.   
We found that secondaries with $q<0.025$ in our fiducial $H/r=0.1$ disk have a next to negligible effect.
In Appendix \ref{app:hr_03} we also present results from a $H/r=0.03$ disk with a high accretion efficiency (i.e., a disk with a high inferred $\alpha$ parameter due to high magnetization) and a $q=0.1$ secondary.
We do not include the results of that simulation in the main text as it was run for only a short amount of time, less well resolved than our $H/r=0.1$ simulations, and likely in a transient state (as argued in Appendix \ref{app:hr_03}).

\section{Results}
\label{sec:results}

To illustrate the main dynamics of the system, in Figure \ref{fig:3D_render} we plot volume renderings of our $q=0.1$ simulation at four different times.
In these renderings the purple/white colors represent cold/hot gas while green represents highly magnetized gas with $\sigma = b^2/\rho=1$ (i.e., the highly magnetized, out-flowing polar regions), where $b^2$ is twice the magnetic pressure in the co-moving frame.
The recent and near-future orbital trajectory of the secondary (i.e., less than one full orbit backwards and forwards in time from the point of the snapshot shown) is denoted by the yellow lines.
Note that only the inner $r<100 \rg$ is rendered in order to avoid obscuration by the outer disk.
The top left panel shows the thin disk around the primary just before the secondary is introduced.
As described in Appendix \ref{app:initial_condition}, this disk is in a magnetically truncated state (c.f.\ \citealt{Liska2022}): the inner 10--15$\rg$ is saturated with vertical magnetic flux with a relatively low $\beta\sim1$, while the bulk of the disk contains turbulent magnetic fields with $\beta \gtrsim 100$ (here $\beta$ is the ratio between thermal and magnetic pressure).
The top right, bottom left, and bottom right panels show how the structure of the disk changes after 4, 12, and 17 impacts of the secondary, respectively.
At these later times the disk is populated by hotter gas and contains spiral shock waves generated by the secondary. 
Each impact also produces plumes of hot gas that escape from both sides of the disk.

Rarely (twice during the duration of the simulation) the secondary produces its own jet of highly magnetized ($\sigma>1$), collimated outflow as seen in the bottom right panel of Figure \ref{fig:3D_render}.
This happens even though the secondary is nonspinning because its motion produces an effective ergosphere and can power jets in an analogous way to the \citet{BZ1977} mechanism for spinning black holes \citep{Neilsen2011,Penna2015,Cayuso2019}.
Effectively, the magnetic fields can extract the kinetic energy of the black hole.
The jet is short lived, persisting for $\sim 600 M$, and has a relatively insignificant power.
Its electromagnetic power (i.e., the Poynting flux integrated over a sphere surrounding the black hole) is a factor of $\gtrsim10$ less than the electromagnetic power associated with the highly magnetized outflow in the polar regions of the primary, which is itself $\lesssim1\%$ of $\dot M c^2$, where $\dot M$ is the accretion rate onto the primary.
Generally, the electromagnetic power in outflows caused by the secondary's interaction with the disk and corona are much larger than the power in the short-lived secondary jets.
The former can also temporarily exceed the power in the $\sigma>1$ outflow around the primary.
The mechanical power in outflows is typically larger than all of these and dominated by gas released by the secondary and coronal winds. 
The reason the secondary's jet appears so prominent in the bottom right panel of Figure \ref{fig:3D_render} despite its weak power is partially an artifact of the rendering process, where we highlight all regions of $\sigma>1$ equally regardless of absolute strength of the magnetic field, but also because it occurs at large distances ($\gtrsim 70$ $\rg$) from the primary in between the polar regions and the corona, where both the magnetic field and mass density are weakest. 
When the secondary reaches this region while still having a non-negligible amount of accumulated magnetic flux (i.e., after crossing the disk but before crossing the pole), a jet is produced.
This happens only twice during our simulation as seen in Figure \ref{fig:orbits}, in particular, the orbits labeled 4 and 8.
At all other times, the dense surrounding matter and strong surrounding magnetic fields suppress any possible jet formation.

\begin{figure*}
\includegraphics[width=0.97\textwidth]{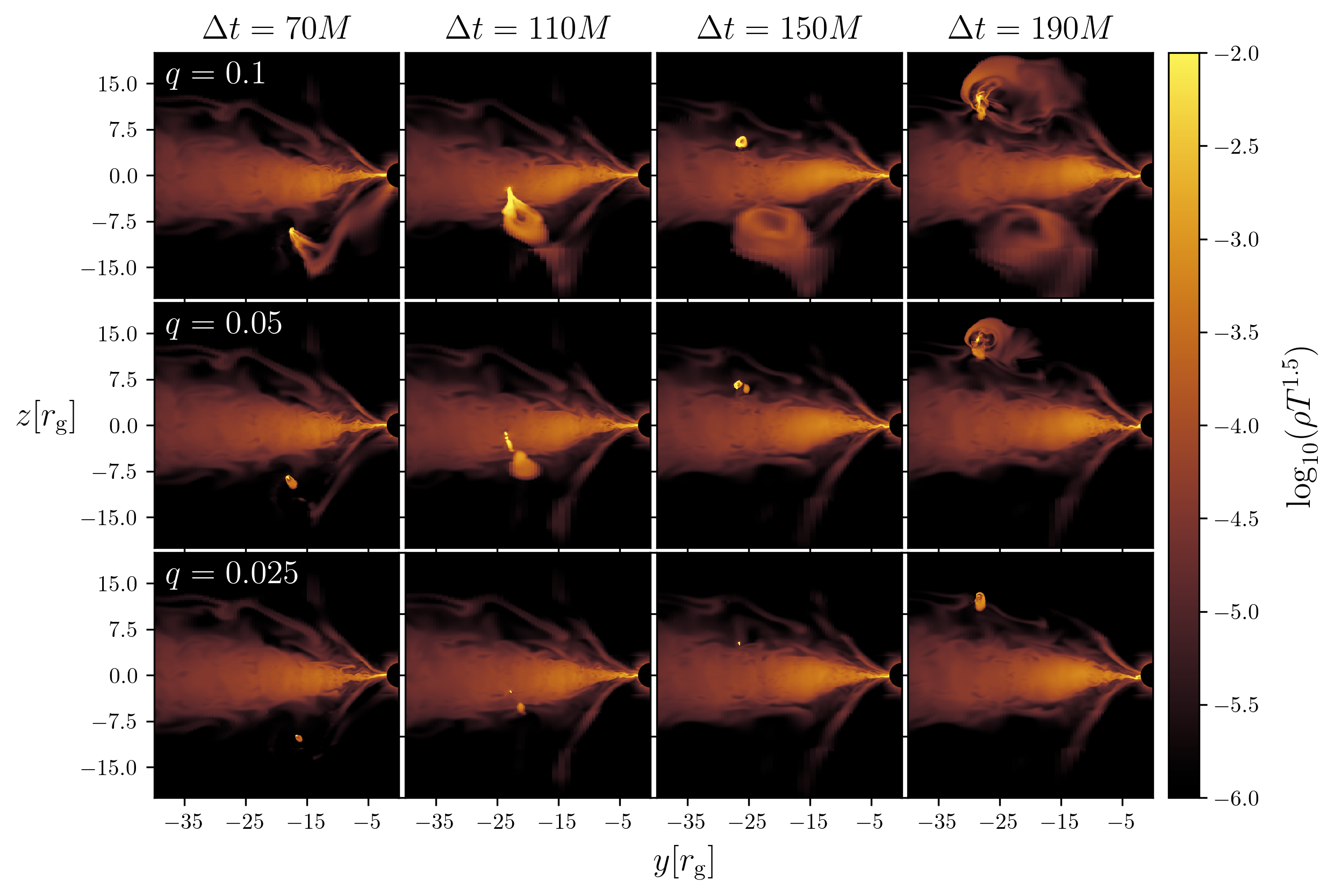}
\caption{Time series of the first impact of the secondary with the primary's accretion disk.
The colormap highlights regions with high $\rho T^{1.5}$, which is a proxy for regions of high emission with our cooling function (Equation \ref{eq:lambda_cool}).
The different rows represent $q=0.1$ (top), $q=0.005$ (middle), and $q=0.0025$ (bottom).  
All three simulations generally show the same behavior, with a shock cone developing around the secondary as is passes through the disk, resulting in two hot gas bubbles expanding on either side.
The sizes of these features decrease with decreasing $q$.
}
\label{fig:impact_timeseries}
\end{figure*}

To isolate the dynamics of the hot plumes seen in Figure \ref{fig:3D_render}, Figure \ref{fig:impact_timeseries} shows a time series of $\rho T^{1.5}$ around the first impact of the secondary with the disk in each of our simulations.  
Because $\rho T^{1.5}$ is roughly proportional to our cooling function (Equation \ref{eq:lambda_cool}), Figure \ref{fig:impact_timeseries} particularly highlights the regions that are contributing to the volume-integrated cooling rate.  
The simulations with different $q$ qualitatively display the same evolution.
The secondary first starts interacting with the corona (i.e., the hot dilute, low $\beta$ region above and below the disk), forming a supersonic bow shock approaching the midplane.  
It then impacts the disk and passes through, creating a bubble of hot gas that expands adiabatically outward from the initial impact point. 
Finally, it leaves the disk, creating another bubble of hot, expanding gas.  
The relative size of the bow shock and the expanding bubbles at a fixed time is $\tilde \propto$ $q$ in each direction, so that while the effects produced by the $q=0.1$ secondary in Figure \ref{fig:impact_timeseries} are dramatic, the effects of the $q=0.025$ secondary are almost unnoticeable, with the effects of the $q=0.05$ secondary somewhere in between. 

To more precisely quantify the energy being added to the gas by the secondary, we calculate the integrated cooling rate as
\begin{equation}
\label{eq:Lcool}
L_{\rm cool} = \int \int \int 
-\Lambda_{\rm cool} u_t \sqrt{-g} dx dy dz,
\end{equation}
where $g$ is the determinant of the metric and the integration is performed over all regions of the simulation that are not within either black hole's event horizon.
$L_{\rm cool}$ calculated in this way is effectively a measurement of the total heating rate and approximately independent of the precise form of $\Lambda_{\rm cool}$.
This is because the cooling times (calculated as $u_{\rm g}/\Lambda_{\rm cool}$, where $u_{\rm g}$ is the internal energy of the gas) in the regions contributing the most to the overall integral in Equation \eqref{eq:Lcool} are short compared to both the crossing time of the secondary through the disk and the viscous heating time of the disk due to MRI turbulence (see Appendix \ref{app:cooling_function} for a discussion).

\begin{figure*}
\includegraphics[width=0.97\textwidth]{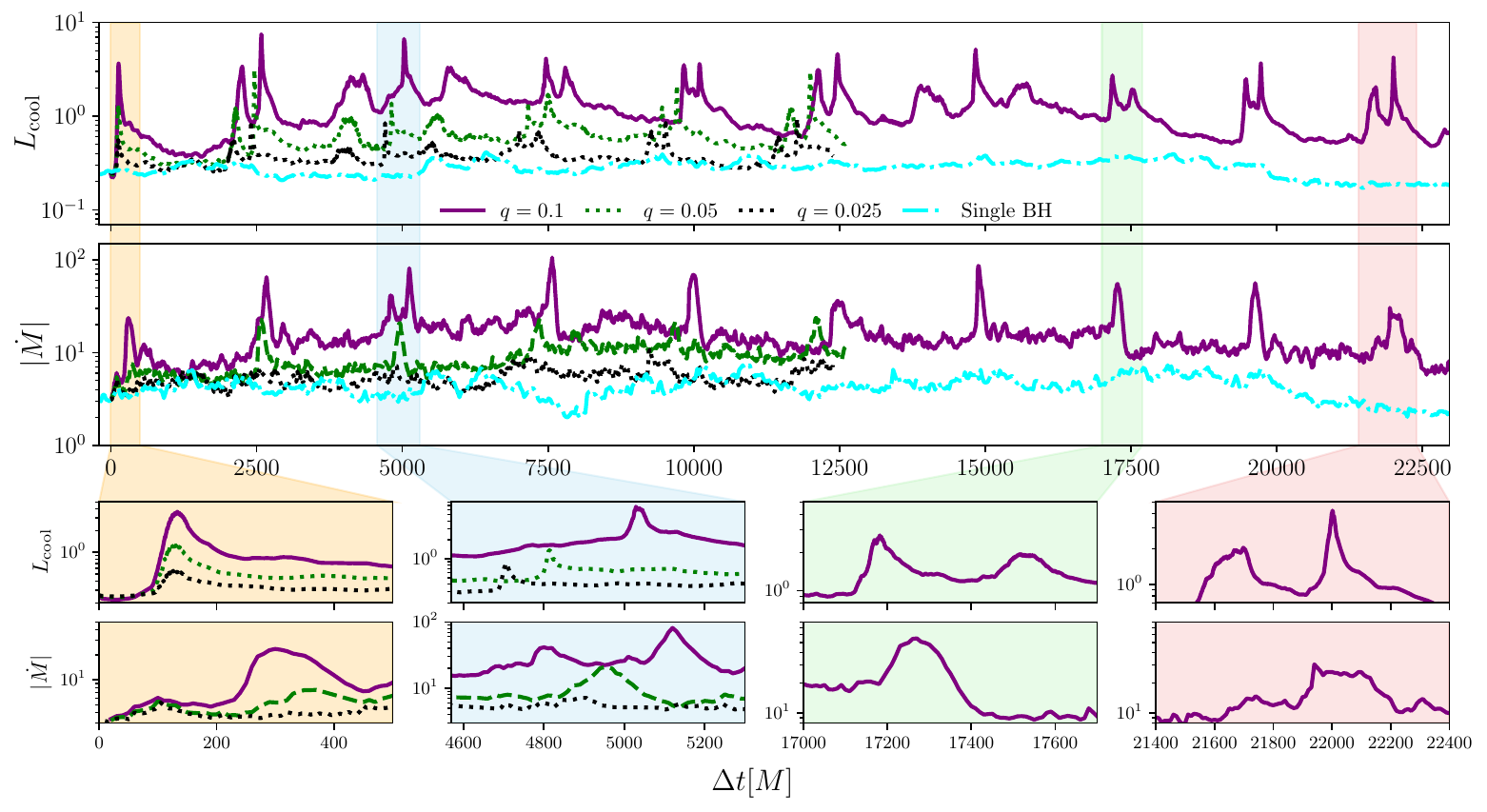}
\caption{Volume-integrated cooling rate, $L_{\rm cool}$, (top panel) and accretion rate, $\dot M$, (second panel) vs.\ time for our three binary simulations compared with the single black hole simulation (cyan dash-dotted line).
The four pairs of panels at the bottom of the figure show zoomed-in views of four disk impact events.
Here $L_{\rm cool}$ is effectively a measure of the instantaneous total heating rate. 
Each impact of the secondary with the disk results in a spike in $L_{\rm cool}$, with broader peaks generally associated with impacts at larger distances from the primary. 
Typically the peaks in $L_{\rm cool}$ are followed by broader peaks in the accretion rate 100--200 $M$ later associated with shocked gas that has lost angular momentum. 
}
\label{fig:Lcool_vt}
\end{figure*}

We plot $L_{\rm cool}$ vs.\ time for each simulation in the top panel of Figure \ref{fig:Lcool_vt} compared with the same quantity in the single black hole simulation.
The simulations that include a secondary black hole show distinct peaks associated with secondary impacts, displaying magnitudes ranging from $\lesssim 12$ times the ``quiescent'' emission (i.e., the emission between secondary impacts) for $q=0.1$ to $\lesssim 3$ times the ``quiescent'' emission for $q=0.025$, though there is a wide range for these values in each simulation.
These peaks occur at slightly different times for each $q$ due to the mild dependence of the orbital timing on the mass of the secondary as discussed in \S \ref{sec:methods}.  
Because the cooling time is short (see Appendix \ref{app:cooling_function}), the peaks occur simultaneously with disk impacts and their widths are determined by the crossing time of the secondary through the disk.
The latter time is different for each peak because the impacts happen at different distances from the black hole where both the disk scale height and secondary speeds are different.
The dependence of the magnitude of individual peaks on $q$ is shallower than the expected $L_{\rm cool} \propto q^2$, roughly $L_{\rm cool}\ \tilde \propto\ q^{1.5}$. 

Akin to the typical variability patterns in the optical light-curves of OJ 287, there are several narrow, double-peaked structures in the top panel of Figure \ref{fig:Lcool_vt}, associated with the rapid motion of the secondary black hole through its periapsis.
Double peaks occur when the periapsis is located outside the body of the disk so that the secondary crosses through one part of the disk and then another in rapid succession at small distances $(\lesssim 25 \rg)$.  
This can be seen by comparing with the bottom panel of Figure \ref{fig:orbits}, showing that double peaks occur when the distance between points where $z_{\rm orbit}$ crosses zero becomes small.
In addition to the narrow double peaks that occur in quick succession (typically happening within a few 100 $M$ of each other), there are also triple-peaked structures (e.g., at $\sim$ 5000 and 15000 $M$) with a sharp narrow peak centered between two weaker and broader peaks separated by longer times (typically within $\sim$ 800 $M$ of each other).  
Triple peaks occur when the periapsis is located within the accretion disk (e.g., between orbits labeled 2--3 and 6--7 in Figure \ref{fig:orbits}).
When this happens, the secondary only crosses the disk once at radii $\lesssim 25\rg$ during its orbit, producing the centered narrow peak.
The broader peaks surrounding this peak are caused by crossings at larger distances.
Note that the secondary jets appear in the simulations $\lesssim$ half an orbital period after the first double peak in the light curve that follows a triple peak (i.e., $\sim$ 1.5 orbital periods after the triple peak) because that is when the orbit precesses to the configuration favorable for jet formation (orbits labeled 4 and 8 in Figure \ref{fig:orbits}) as discussed previously in this section.  

In between each peak, the quiescent luminosity of the disk also depends on $q$.
This is because not only does the secondary accumulate an accretion flow itself of hot, shocked gas, but it can also heat up the primary accretion disk through the repeated impacts.
Both of these increase the overall quiescent emission, but the heating of the disk by repeated impacts is the dominant effect.

In the second panel of Figure \ref{fig:Lcool_vt}, we plot the accretion rate vs.\ time for each of the simulations on the same time axis as the volume-integrated cooling rate, while in the bottom panel we show zoomed-in plots of both $L_{\rm cool}$ and $\dot M$ for four different impact events.  
Generally, passages through the disk by the secondary are associated with spikes in the accretion rate.
These spikes, however, are typically broader than those in $L_{\rm cool}$ and come after a time delay of 100--200 $M$.
They are typically associated with shocked gas from the disk that has lost a significant amount of angular momentum and quickly falls inwards towards the primary black hole.  
Because the infall time of this gas is short compared to its cooling time, there are no secondary peaks in $L_{\rm cool}$ caused by these accretion events, although this may be an artifact of the ad-hoc nature of our cooling function.  
The quiescent accretion rate is also increased by the secondary, which can significantly alter the structure of the disk and drive material inwards due to enhanced angular momentum transport from non-axisymmetric shocks.

\begin{figure*}
\includegraphics[width=0.97\textwidth]{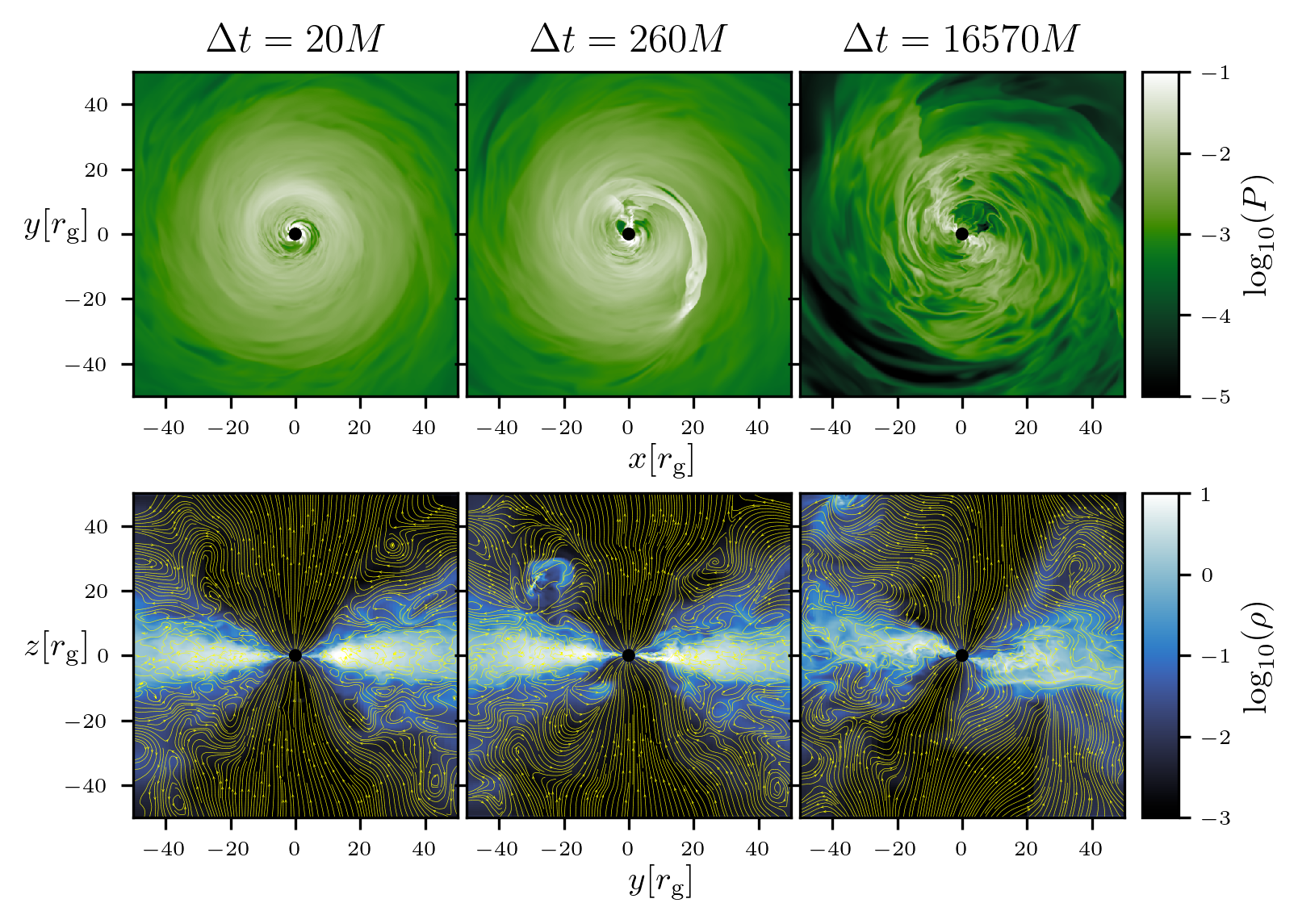}
\caption{Effects of the secondary on the primary accretion disk for $q=0.1$ shown in the midplane pressure (top panel) and a poloidal slice of mass density with magnetic field lines over-plotted (bottom panel).  
The first column is right after the secondary has been added to the simulation, the second column is just after the first impact of the secondary, and the third column is later in the simulation after $\sim15$ impacts. 
Each impact produces a spiral shock wave that temporarily boosts the accretion rate (compare with the bottom panel of Figure \ref{fig:Lcool_vt}). 
The repeated interaction with the gravity of the secondary also torques the disk in non-trivial ways, resulting in tilts in both the $y$--$z$ and $x$--$z$ plane. 
Animations of this figure are available for the $q=0.1$ (\url{https://youtu.be/v1-WId2JKbc}), $q=0.05$ (\url{https://youtu.be/3aIjSRte-eM}), and $q=0.025$ (\url{https://youtu.be/6qrMa5xgzpg}) simulations.}
\label{fig:midplane_timeseries}
\end{figure*}

To more clearly see the effects of the secondary over time, we highlight the pressure and mass density in our $q=0.1$ simulation at three different times in Figure \ref{fig:midplane_timeseries}.
In particular, the first column shows an instance right after the secondary is introduced, the second column shows an instance just after the first impact of the secondary with the disk, and the third panel shows an instance after $\sim$ 15 impacts of the secondary.  
The top row shows midplane slices of pressure.
The middle panel of this row demonstrates how individual impacts lead to the enhanced accretion events seen in Figure \ref{fig:Lcool_vt} through transient spiral shock structures.
After many of these impacts, the repeated propagation of these shock waves through the disk completely alters the midplane structure, as seen in the right panel of Figure \ref{fig:midplane_timeseries}.
In particular, the gas is significantly more turbulent throughout and the accretion within the ISCO radius no longer proceeds through well defined inflowing spiral gas streams.
This is because the gas can have non-negligible angular momentum in directions other than the $z$-direction (the initial angular momentum direction), which is evident in the $y$-$z$ slices of mass density overplotted with magnetic field lines in the bottom row of Figure \ref{fig:midplane_timeseries}.
The right panel shows that the disk and the magnetic field has significantly tilted towards the positive $y$ axis by $15$--$20\degree$.  
In the $x$-$z$ plane (not shown) the disk and magnetic field have also tilted by a similar amount towards the negative $x$ axis.  
The field geometry is also significantly distorted near the secondary as it drags magnetic field lines from both the disk and corona.  

\begin{figure}
\includegraphics[width=0.47\textwidth]{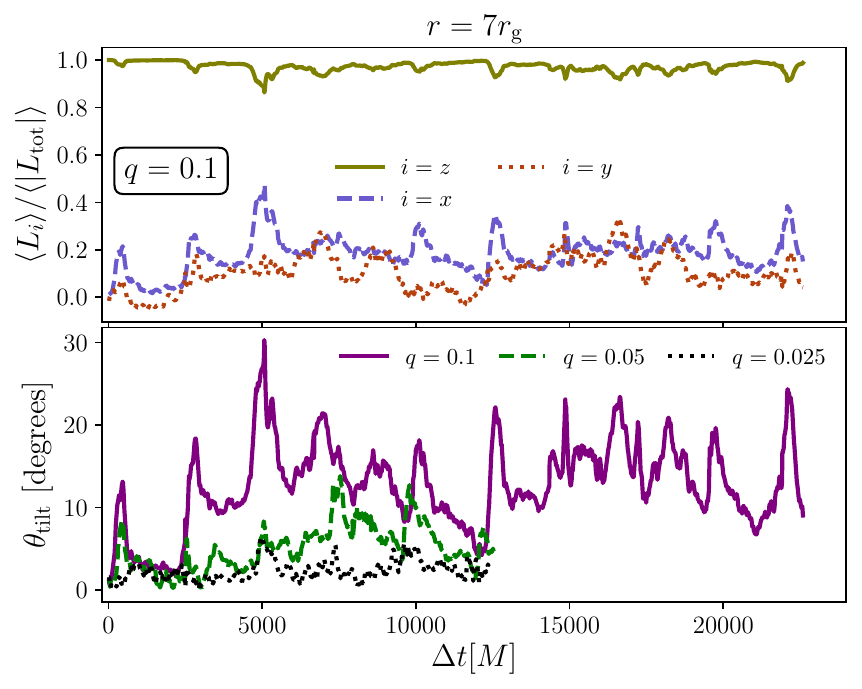}
\caption{
Angular momentum direction vs.\ time at $r=7\rg$ in our simulations.
Top: angle-averaged $x$, $y$, and $z$ components of angular momentum, $L_i$, relative to the total angular momentum, $l_{\rm tot}$, in our $q=0.1$ simulation.
Bottom: Tilt angle, $\theta_{\rm tilt} = \arccos{(\langle L_z\rangle/\langle L_{\rm tot} \rangle)}$ in our $q=0.1$ (solid purple), $q=0.05$ (dotted green), and $q=0.025$ (dotted black) simulations.
Gravitational torque from the secondary black hole can cause significant tilting of the disk (10--20$\degree$ for $q=0.1$, 5--10$\degree$ for $q=0.05$, and a few degrees for $q=0.0025$) in both the $x$ and $y$ directions (note that the secondary's orbit is in the $y$-$z$ plane).
This tilt, however, does not strictly increase with time but seems to vary around a saturated value.
Spikes in the tilt angle are seen during increased accretion events associated with lower angular momentum spiral streams caused by the secondary's impacts with the primary accretion disk (see Figures \ref{fig:Lcool_vt} and \ref{fig:midplane_timeseries}).
}
\label{fig:angular_momentum_vt}
\end{figure}

To quantify the disk tilt more precisely, in the top panel of Figure \ref{fig:angular_momentum_vt} we plot the time evolution of the angular momentum components ($L_i$ for $i\in [x,y,z])$ of the disk relative to the total angular momentum ($L_{\rm tot}$) at $r=7 \rg$ for the $q=0.1$ simulation. 
Both the $x$ and $y$ components of the angular momentum increase from initially zero to $L_x/|L_{\rm tot}|,L_y/|L_{\rm tot}| \sim 0.2 $ and then tend to slowly vary around that value.  
There are distinct peaks in $L_x/|L_{\rm tot}|$ that coincide with the peaks in accretion rate seen in the bottom panel of Figure \ref{fig:Lcool_vt}.  
This is because the lower angular momentum streams shocked by the secondary temporarily increase the local measurement of $L_x/|L_{\rm tot}|$.
Otherwise, there seems to be no general increase or secular trend of the relative $x$ and $y$ component of angular momentum.  
More precisely, we define the tilt angle as $\theta_{\rm tilt} = \arccos (L_{z}/|L_{\rm tot}|)$ and plot it vs.\ time in the bottom panel of Figure \ref{fig:angular_momentum_vt} for all of our simulations.
Other than the peaks, the $q=0.1$ disk tends to oscillate around $\theta_{\rm tilt} \sim 10$--20$\degree$, while the $q=0.05$ disk tends to oscillate around $\theta_{\rm tilt} \sim $5--10$\degree$ and the $q=0.025$ disk tends to oscillate around $\theta_{\rm tilt} \sim$ a few degrees.

As we discuss in Appendix \ref{app:torques}, this tilting of the accretion disk is caused by the gravitational torque of the secondary.  
This happens in a highly nonlinear regime because the instantaneous torques at different points in the secondary's orbit can be quite large \citep{Ivanov2024}, unlike the case of secondary objects orbiting at large distances (e.g., \citealt{Tremaine2014}).
As a result, the angular momentum vector of the disk does not simply precess or monotonically align with the orbital plane of the secondary but varies stochastically around a saturated value.  
It is unclear whether this behavior will continue on longer time-scales or if the disk will ultimately align with the orbit plane.

\section{Towards More Realistic Simulations of OJ 287}
\label{sec:limitations}
Our simulations have two evident limitations: (a) we do not include proper photon radiation transport, and b) the disk thickness in our models is likely larger than that in OJ 287.  

The main disadvantage of a) is that we cannot generate realistic light curves. 
Instead, our measurements of the volume-integrated cooling rate correspond with the instantaneous intrinsic heating rate, including that of both the quiescent disk and the secondary impacts. 
In reality, however, the energy added to the system by the impacts is not likely to be radiated immediately but gets trapped in optically thick bubbles that cool through adiabatic expansion.  
Only once the bubbles become sufficiently optically thin do they release this energy.  
This means that not only will the flares be slightly delayed from what we see in our simulations, but their duration could be different because the relevant time scale is the expansion time and not the disk-crossing time.
Furthermore, including the dynamical effects of radiation could alter the thermodynamic properties of the disk. 
For the accretion rate regime relevant to our study, radiation pressure likely dominates over gas and magnetic pressures for most regions of the disk (e.g., \citealt{Shakura1973,Valtonen2019}).  
On the other hand, \citet{Zhang2025} found that in 3D radiative GRMHD simulations, even when the radiation pressure dominates in magnitude the disk can still be gas pressure supported because its \emph{gradient} is larger (note, however, that their simulations were for stellar mass black holes).
This limitation can be overcome in the future with readily available computational tools that solve the full radiative GRMHD equations (e.g., \citealt{Anninos2005,Fragile2014,McKinney2014,Sadowski2014,Sadowski2016,BHLIGHT,Liska2022,Mishra2022,White2023,Zhang2025,Zhang2025b}).

There is an additional complication, however, in that the energy removed from our simulations is ``gray,'' that is, it is not partitioned into different photon energy bands.  In reality the bulk of the disk emission would peak at a lower photon frequency than flare emission (which for OJ 287 is in the optical band).  
As a result, the flares would stand out more clearly against the background of disk emission than they do in our simulations, where we only see clear peaks when the total power radiated from a disk impact is much higher than the power radiated by the disk.
To properly model the frequency dependent radiation, post-processing GR ray tracing \citep{Noble2007,grtrans,raptor1, ipole,White2022} or Monte Carlo \citep{Dolence2009,kappamonty} calculations will be required. 


The main disadvantage of b) is that our results must be extrapolated to more realistic parameters for OJ 287 (and other systems).
We argued in \S \ref{sec:analytics} how this could be done, namely by identifying the pairs of $q,H/r$ that correspond to the same physical regime as the observed systems (for a fixed effective viscosity $\alpha$).
For instance, the analytic expectation is that in terms of flare energetics, the relative importance of the secondary is expected to scale as $q^2/(H/r)^3$, while in terms of gravitational torques the relative importance of the secondary is expected to scale as $q/(H/r)^2$.
So if OJ 287's disk has $H/r=0.03$, this would argue that our $q=0.05$ simulations best represent the flare dynamics while our $q=0.1$ simulations best represent the tilting dynamics. 
These scaling relations, however, are based on an $\alpha$-disk model that cannot fully account for the nonlinear GRMHD dynamics of the system.  
For instance, we find that $Q_{\rm s}$, the integrated heating rate induced by the secondary during a disk impact as measured by our integrated cooling rate $L_{\rm cool}$ (Figure \ref{fig:Lcool_vt}), scales more like $q^{1.5}$ than the expected $q^2$.
Furthermore, as we show in Appendix \ref{app:hr_03}, if the thinner disk has a much higher accretion efficiency ($\alpha$) than the thicker disk, the impact power of the secondary can actually be less relatively significant, which is contrary to a naive extrapolation. 
Systematically exploring the regime of smaller $H/r$ and smaller $q$ will require higher resolution simulations enabled by GPU acceleration  (e.g., \citealt{HAMR,Stone2024, Zhang2025}).

In addition to a) and b), we have also neglected black hole spin for both the primary and secondary black holes. 
The precessing binary model for OJ 287 constrains the primary spin to be $\chi \sim 0.3$ \citep{Valtonen2016}, while the spin for the secondary is relatively unconstrained.
Nonzero spin of the primary could induce precession or further tilt of the accretion disk as the spin axis precesses due to spin-orbit coupling \citep{Ressler2024}; note however, that this is a small effect for OJ 287's parameters in the standard model because the mass ratio is so small. 
Spinning black holes could also produce strong jets, and the interaction between these jets could cause additional flaring behavior \citep{Palenzuela2010,Gutierrez2024,Ressler2025,Ennoggi2025}.
Moreover, the secondary could either indirectly (via perturbations to the disk) or directly (via gravitational effects on polar magnetic field lines) induce significant variability in the primary jet position angle \citep{Agudo2012,Traianou2025}.

Finally, we note that realistic initial/boundary conditions for the accretion disk are not currently known.
This is true not only for OJ 287 but for most AGN in general.
Here, we focused on one particular realization of the primary accretion flow in which the inner disk (10--15 $\rg$) is magnetically truncated by an accumulation of vertical magnetic field while the outer parts of the disk are weakly magnetized.
Other possibilities include magnetically arrested thin disks \citep{Avara2016,Scepi2024,Most2024,Wang2025b}, ultra-magnetized disks \citep{Hopkins2024,Kaaz2025,Guo2025b,Wang2025}, disks in the ``standard and normal evolution'' (SANE) regime where the magnetic field dynamics are dominated by MRI turbulence (e.g., \citealt{Noble2009,Noble2010,Noble2012,Zhang2025}), magnetically dominated ``ergomagnetospheres'' \citep{Blandford2022}, or another yet undiscovered state determined by more realistic galactic-scale feeding.
How sensitive our results are to the accretion state of the primary flow is unclear without a wider parameter survey.  


\section{Conclusions}
\label{sec:conc}
We have presented global 3D GRMHD simulations of an OJ 287-like system comprised of a secondary black hole plunging through the accretion disk of a much larger black hole.  
In particular, expanding on our previous work on thick disks \citep{Ressler2024}, we have considered a thin, $H/r=0.1$ primary disk.
Contrary to the thick disk case, we found that secondaries with $q \gtrsim 0.05$ can have a significant impact on the accretion flow (Figure \ref{fig:3D_render}) and produce highly energetic impact events.
These events are associated with  supersonic shocks that pass through the disk and create hot, expanding bubbles on both sides (Figure \ref{fig:impact_timeseries}).
The shocks themselves dissipate energy at a rate that can well exceed the ``quiescent'' heating rate of the entire disk during the secondary crossing as measured by the total volume-integrated cooling rate (Figure \ref{fig:Lcool_vt}).
Because in reality most of the radiation associated with these events would be radiated at higher photon frequencies than the bulk of the disk emission, even our smallest secondary of $q=0.025$ (where the volume-integrated cooling rate increases only by a factor of $\lesssim2$ during the brief periods of enhanced heating) would likely produce high-contrast flares in, e.g., the optical band for OJ 287. 

Dynamically, the secondary impact events produce spiral shocks (Figure \ref{fig:midplane_timeseries}) that lead to enhanced accretion events (compare with Figure \ref{fig:Lcool_vt}) that are delayed by a few 100s of $M$ from disk impacts.
Over time these shocks significantly modify the disk structure and increase the quiescent volume-integrated cooling rate from the unperturbed disk value by $\sim$ 10\% for $q=0.025$ and up to a factor of $\sim$ 4 for $q=0.1$.
Additionally, as the secondary passes through its precessing orbit it exerts significant instantaneous net torques on the accretion disk, causing it to tilt from its initial alignment with the $+z$ axis (Figure \ref{fig:angular_momentum_vt}).
Because this interaction is nonlinear (see also the discussion in Appendix \ref{app:torques}), the time variability of the disk tilt angle in this regime is stochastic around a saturated value.
Specifically, we do not see evidence for simple precession or monotonic alignment with the orbital plane of the secondary. 
Instead, we find that the tilt angle of the disk varies between 10--20$\degree$ for $q=0.1$, 5--10$\degree$ for $q=0.05$, and 0--5$\degree$ for $q=0.025$.

We find that the motion of secondary can produce small, electromagnetically dominated jets (bottom right panel of Figure \ref{fig:3D_render}) roughly once every 4 orbits, which would correspond to every $\sim$ 48 years for OJ 287.
The power in these jets, however, is substantially weaker than all the other outflows in the simulation and likely not dynamically or observationally significant even for our largest mass ratio $q=0.1$.
If the secondary were rapidly spinning, however, its jet power could be as much as $\sim$ 10--100 times larger [i.e., by a factor of $(\chi_{\rm s}^2 + v_{\rm orbit}^2)/v_{\rm orbit}^2$, where $\chi_{\rm s}$ is the dimensionless spin of the secondary, see \citealt{BZ1977,Sasha2011,Neilsen2011,Penna2015}], and if it were able to accumulate a larger amount of magnetic flux than we see our simulations (e.g., if the primary disk was more magnetized) it could be further increased by a similar factor.\footnote{This is because the jet power scales as $\phi_{\rm BH}^2$, where $\phi_{\rm BH} = \Phi_{\rm BH}/\sqrt{|\dot M|}$ and $\Phi_{\rm BH}$ is the magnetic flux threading the black hole. 
GRMHD simulations of thin disks have found that there is a maximum value of $\phi_{\rm BH}$ at $\sim$ 50 \citep{Avara2016,Scepi2024}, while our secondary has only a modest $\phi_{\rm BH}\sim$ a few.}
If so, this jet could provide an additional flaring mechanism that would be four times as rare as disk-impact flares and occur several years after the expected collisions (although the duration and timing of the flares likely also depend on jet power).  
Nevertheless, using our current simulations as a guide, only in the most optimistic scenario is this increased power large enough to possibly produce observable flares for $q=0.1$. 
Jets from smaller $q$ secondaries interacting are likely energetically unimportant for  $H/r=0.1$ disks but could become relevant for smaller $H/r$ disks.\footnote{A simple estimate of the ratio between the secondary's jet power and the overall luminosity of the disk is $\tilde \propto\   |\dot M_{\rm s}/\dot M| \propto \alpha^{-1} q^2/(H/r)^3$, where $\dot M_{\rm s}$ is the accretion rate onto the secondary (which we take to be the Bondi-Hoyle-Lyttleton accretion rate) and we have assumed that the jet power is proportional to the accretion rate. } 
Simulations that include secondary black hole spin and/or have more strongly magnetized primary disks will be required to investigate this possibility.

The temporal properties of the ``flares'' in our simulations are consistent with the observed optical variability of OJ 287 in that they are almost all characterized by double-peaked structures (Figure \ref{fig:Lcool_vt}) resulting from back-to-back disk cotillions near the periapsis of the secondary. 
This is expected since we assume a similar orbit as that in the standard precessing binary model.
Qualitatively, we also observe several processes in our simulations that have been invoked to explain specific variability patterns, such as delayed accretion events \citep{Valtonen2017,Komossa2020,MOMO4}, spiral shock waves \citep{Pihajoki2013}, bending of the accretion disk due to the gravitational torque of the secondary \citep{Valtonen2007,Valtonen2023}, and secondary jets \citep{Pihajoki2013b,Valtonen2024b}.
Determining whether or not these are viable explanations of the data will require more realistic simulations with accurate modeling of electromagnetic radiation as discussed in \S \ref{sec:limitations}.

Our previous work on thick disks \citep{Ressler2024} found that even for $q=0.1$, the impacts of the secondary on the disk had a negligible effect on the accretion dynamics, in stark contrast to our findings here.  
This is likely for two reasons.  
First, the gas in thick disks is nearly virial, meaning that the sound speed is comparable to the orbital speed.  
Because of this, the shocks from the secondary have Mach numbers $\lesssim$ 1 and the gas is only heated to temperatures modestly above the average temperature at the orbital radius.
Second, the inflow velocity in thick disks is much faster than in thin disks due to both a larger effective viscosity and larger scale height.  
Consequently, the disk can rapidly respond to changes induced by the secondary (or in other words, the typical turbulent motions of the gas are comparable to or larger than the secondary-induced perturbations).
On the other had, in \citet{Ressler2024} we found that spin-orbit coupling between the primary black hole's spin and the orbital angular momentum of the secondary could cause significant jet precession.
In this work, since we have neglected black hole spin for simplicity, we can only speculate whether the jet in thin disk simulations would behave in the same way.
For instance, while highly magnetized thick disk (like the ones we studied in \citealt{Ressler2024}) are particularly efficient at aligning the disk and jet with the spin axis of the central black hole \citep{McKinney2013,Ressler2023,Chatterjee2025}, thinner disks with weaker magnetization can break and/or precess when misaligned with the spin axis \citep{Teixeira2014,Liska2020,Kaaz2023b}.
How the jet from a thin disk would respond to a time-variable, misaligned spin axis, and whether this can explain, e.g., the ribbon-like morphology of the OJ 287 jet \citep{Traianou2025} is a key question for future work to answer. 

While we have focused here on only a single disk scale height ratio, $H/r=0.1$, and three binary mass ratios, $q\in (0.1,0.05,0.025)$, the magnetohydrodynamics and thermodynamic results are still highly relevant for OJ 287-like systems.  
This is because not only are the qualitative effects likely universal for small-mass-ratio binaries with thin disks, we can plausibly (but cautiously) extrapolate the relative importance of these effects to different parameters using simple analytic arguments (see \S \ref{sec:analytics}). 
That said, our limited set of simulations are no substitute for a full parameter survey or a simulation with the precise constrained parameters of OJ 287 in the precessing binary model.  
This is especially true given the uncertainties in the dynamics of typical AGN disks (see Appendix \ref{app:hr_03} for a discussion).

Indeed, our simulations represent only a first step towards realistic, global simulations of OJ 287 and similar systems. 
More detailed comparison with observations will require accurate modeling of the thermodynamic and dynamical effects of radiation, time-dependent GR ray tracing/Monte Carlo post-processing emission calculations, and higher resolution to study thinner disks that better represent typical AGN (as well as smaller secondary black holes).
All of these additions are now feasible with current computational tools and resources.  
We thus expect that continued progress in these directions will yield significant gain for our understanding of the prototypical supermassive black hole binary system OJ 287 and likely other LISA/PTA gravitational wave targets.

\begin{acknowledgments}
We thank L.\ Zhang, J.M.\ Stone, J.\ Rule, N.\ Murray, C.\ Thompson, E.R.\ Most, and R.\ Blandford for useful discussions.
We acknowledge the support of the Natural Sciences and Engineering Research Council of Canada (NSERC), [funding reference number 568580]
Cette recherche a \'et\'e financ\'ee par le Conseil de recherches en sciences naturelles et en g\'enie du Canada (CRSNG), [num\'ero de r\'ef\'erence 568580].
L.C.\ is supported in part by
Perimeter Institute for Theoretical Physics.
Research at Perimeter Institute is supported in part by the Government of Canada through the Department of Innovation, Science and Economic Development Canada and by the Province of Ontario through the Ministry of Colleges and Universities. 
B.R.\ is supported by the Natural Sciences \& Engineering Research Council of Canada (NSERC), the Canadian Space Agency (23JWGO2A01), and by a grant from the Simons Foundation (MP-SCMPS-00001470). B.R. acknowledges a guest researcher position at the Flatiron Institute, supported by the Simons Foundation.
This research was supported in part by grant NSF PHY-2309135 to the Kavli Institute for Theoretical Physics (KITP).

The computational resources and services used in this work were partially provided by facilities supported by the VSC (Flemish Supercomputer Center), funded by the Research Foundation Flanders (FWO) and the Flemish Government – department EWI and by Compute Ontario and the Digital Research Alliance of Canada (alliancecan.ca).

During the preparation of this manuscript, we used the {\tt ChatGPT} software \citep{chatgpt} for suggestions on data analysis, plotting, and minor grammatical editing. 

\end{acknowledgments}

\software{{\tt Athena++} \citep{White2016,Athenapp},
{\tt CBwaves} \citep{cbwaves}, {\tt ChatGPT} \citep{chatgpt},
{\tt matplotlib} \citep{Hunter:2007}, {\tt yt} \citep{yt_code}}

\bibliographystyle{aasjournal}
\bibliography{oj}

\appendix

\section{Single Black Hole Thin Disk Simulation}
\label{app:initial_condition}
\begin{figure*}
\includegraphics[width=0.97\textwidth]{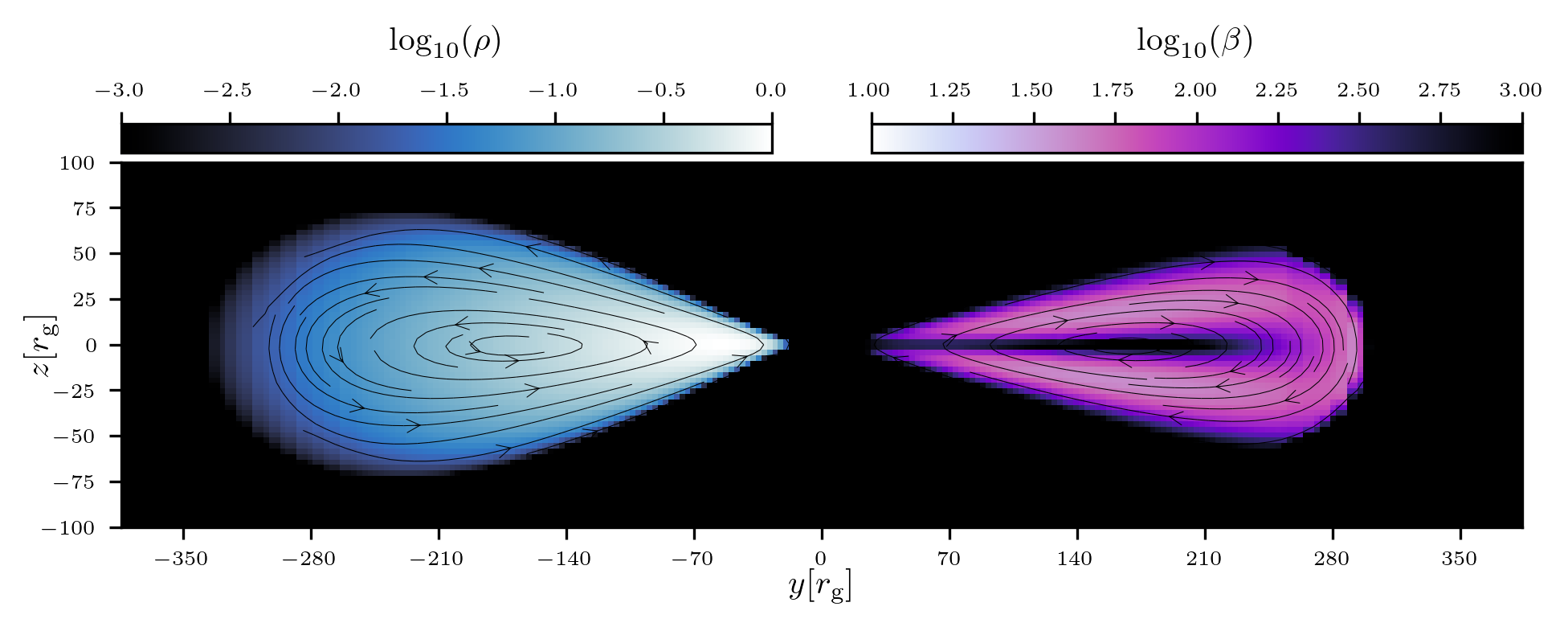}
\caption{The initial conditions for our single black hole, thin disk simulation.  The \citet{Chakrabarti1985} disk is initially in hydrodynamic equilibrium threaded by a single loop of weak magnetic field such that $
\max{(P)}/\max{(b^2/2)} = 100$.  }
\label{fig:initial_conditions}
\end{figure*}
Before adding the binary companion, we initialize a \citet{Chakrabarti1985} disk in hydrodynamic equilibrium around a non-spinning ($\chi=0$) black hole.  
We use a disk with inner edge of $16 \rg$ and pressure maximum located at $53\rg$, with an angular velocity profile of $\Omega = u^\varphi/u^t \propto \lambda^{-n}$ for $n=1.5637$ and $\lambda = \sqrt{-g^{tt}/g^{\varphi \varphi}} $ \citep{Chakrabarti1985,DeVilliers2003}. 
This initial state has $H/r=0.1$ at the pressure maximum when measured as $H/r = \int \rho |\theta-{\rm \pi}/2| d\Omega/\int \rho  d\Omega$, where $d\Omega = \gdet \sin(\theta)d\theta d\varphi$, and $r, \theta,\varphi$ are the spherical Kerr-Schild coordinates converted from Cartesian Kerr-Schild coordinates.
We seed the disk with a single poloidal loop of magnetic field with vector potential 
\begin{equation}
A_\varphi = 0.345 \left(\rho - 0.02\right) r^2 \sin^4(x_\theta) \exp(- r/235),
\end{equation}
where $x_\theta = (\theta - 3{\rm \pi}/8 )/({\rm \pi}/4)$ for $3{\rm \pi}/8<\theta<5{\rm \pi}/8$ and $x_\theta=0$ otherwise.
This vector potential is normalized such that $\max(P)/\max{(b^2/2)} = 100$ in the disk.  
To compute the magnetic field as the curl of the vector potential, $A_\varphi$ is first converted to Cartesian Kerr-Schild. 
We show $\rho$ and $\beta$ for the initial conditions in Figure \ref{fig:initial_conditions} with magnetic field lines overplotted.

We use a computational grid in Cartesian Kerr-Schild coordinates spanning $|x,y,z|
\le 384 \rg$ with base resolution of $96 \times 96 \times 288$ cells.
We then add 7 levels of mesh refinement, particularly focused within one scale height of the midplane as summarized in Table \ref{tab:resolution}.
This grid resolves $H \sim 0.1 r$ by roughly 30--60 cells in the $z$-direction for all radii between $\sim$ $6\rg$ and $80\rg$.  
Within $6\rg$ (i.e., radius of the ISCO), less resolution is required because the gas quickly falls into the black hole without the need for MRI transport.
Outside of $80\rg$ the resolution is about 10--20 cells per scale height in the $z$-direction, and higher resolution is not required because the inflow times are much longer than the length of the simulation.  
The fact that we have $3$ times the number of cells in the $z$ direction as the other two directions in a cubical box means that our cells are elongated in the $x$ and $y$ plane. 
However, similar or more extreme aspect ratios are often used in spherical-polar Kerr-Schild coordinates to better resolve the midplane to good effect \citep{Gammie2003,Noble2009,Noble2010,Porth2019}.  

We use the HLLE Riemann solver \citep{Einfeldt1988} and the piece-wise linear reconstruction method \citep{vanleer1974}.
We set the density and pressure floors to $10^{-7}$ and $10^{-12}$, respectively, while we enforce $\sigma \equiv b^\mu b_\mu/\rho \le 100$ and $\beta \ge 0.001$ with additional density and pressure floors.  
We limit the velocity of the gas such that the maximum bulk Lorentz factor is $<50$.  

\begin{figure}
\includegraphics[width=0.47\textwidth]{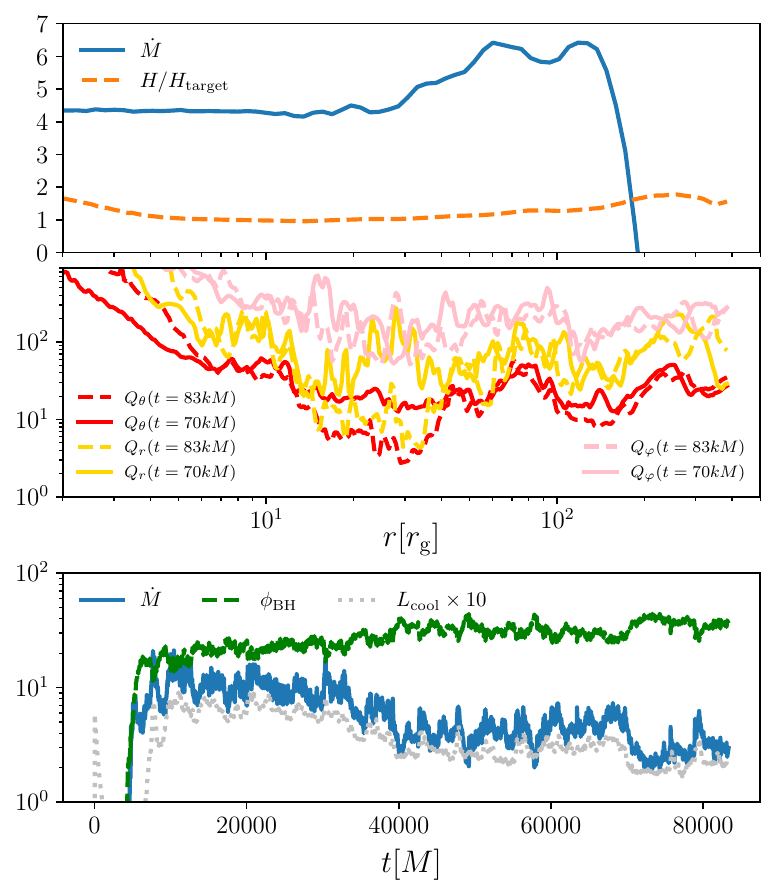}
\caption{Integrated quantities in our single black hole $H/r=0.1$ disk simulation vs.\ radius and time. 
Top: Time-averaged ($40{, }000$--$70{, }000M$) accretion rate, $\dot M$ (solid blue), and the scale height normalized to the target scale height, $H/H_{\rm target}$ (dashed orange) vs.\ radius.  
Middle: MRI quality factors defined around Equation \eqref{eq:mri_lambda}, $Q_{(i)}$, vs.\ radius at $t=70{, }000M$ (solid) and $t=83{, }000M$ (dashed).
Bottom: Accretion rate (solid blue), $\dot M$, dimensionless magnetic flux threading the event horizon (dashed green), $\phi_{\rm BH}$, and volume-integrated cooling rate (dashed silver), $l_{\rm cool}$, vs.\ time.
The simulation reaches inflow equilibrium out to $\sim$ 25$\rg$ (as evidenced by the flat accretion rate profile vs.\ radius) and the cooling function keeps the scale height close to the target except for inside the ISCO radius and at large distances.
The MRI is well resolved in all directions until the quality factors drop after $\sim 70{, }000M$, at which point the accretion rate also drops by a factor of $\sim 2$.  
$\phi_{\rm BH}$ saturates around $\sim 40$, and its variability is almost entirely driven by variability in $\dot M$.
}
\label{fig:1d_torus_plots}
\end{figure}

We evolve the disk for $70,000 M$, at which point the MRI begins to be under-resolved. 
The disk reaches inflow equilibrium out to $\sim$ $25\rg$, as determined from the time-averaged accretion rate shown in the top panel of Figure \ref{fig:1d_torus_plots} alongside the scale height normalized to the target value, $H_{\rm target} =0.1 r$.  
The cooling function effectively keeps $H/H_{\rm target}\sim 1$ outside the ISCO radius (where the flow rapidly falls inwards before it can cool) and within $\lesssim 100 \rg$ (where the cooling time is long compared to the length of the simulation).  

\begin{figure}
\includegraphics[width=0.47\textwidth]{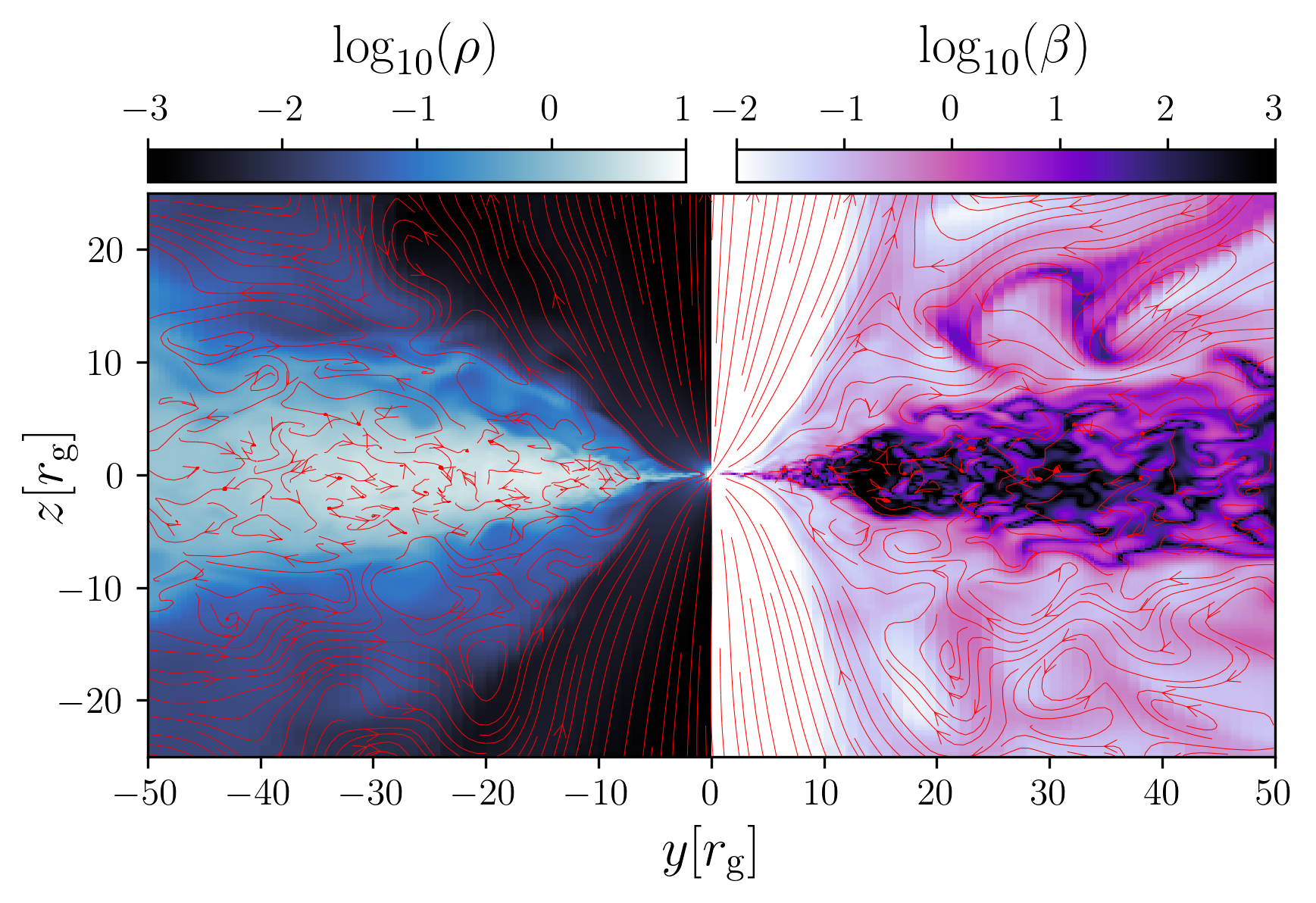}
\caption{Poloidal slice of density, $\rho$, (left) and plasma $\beta$ (right) with magnetic field lines overplotted at $t=50{, }000M$ from our single black hole, thin disk simulation.  The disk is MRI turbulent with $\beta\gtrsim 100$ surrounded by a lower density, highly magnetized corona with $\beta \sim 0.1$.  
Coherent vertical magnetic field populates the inner $10$--$15\rg$ and partially truncates the disk.  }
\label{fig:final_state}
\end{figure}

A poloidal snapshot of density, plasma $\beta$, and magnetic field lines from this simulation is shown at $t=50{, }000 M$ in Figure \ref{fig:final_state}.  
By this time the flow has settled down into a weakly magnetized disk with $\beta\gtrsim 100$ surrounded by a highly magnetized corona with $\beta \sim 0.1$.  
This happens because much of the magnetic field that has been amplified by the MRI has either buoyantly risen out of the main disk body (particularly the non-vertical field) or accumulated in the inner $\lesssim$ 10--15$\rg$ in coherent vertical field lines that bend around the surface of the disk.  
This structure bears a resemblance to recent ``magnetically truncated disk'' simulations \citep{Liska2022}.  
Specifically, the inner 10--15$\rg$ displays a drop in density and has a nearly constant amount of vertical magnetic flux that changes very little after saturation, never displaying the large-amplitude flux eruptions characteristic of thick magnetically arrested disks \citep{Narayan2003,Igumenshchev2003,Sasha2011,Dexter2020b,Ripperda2020,Porth2021,Ripperda2022}.
We also find that a magnetic tower with $\sigma=b^2/\rho>1$ forms in the polar regions that can drive outflow, but the power of this outflow is relatively weak compared to jets associated with highly spinning black holes (since we use $\chi=0$).
A more detailed analysis of the GRMHD dynamics of this state is outside of the focus of this work.  

The time variability of the accretion rate, $\dot M$, dimensionless magnetic flux threading the black hole, $\phi_{\rm BH} \equiv \Phi_{\rm BH}/\sqrt{|\dot M|}$, where $\Phi_{\rm BH}$ is the magnetic flux threading the black hole, and the volume-integrated cooling rate, $L_{\rm cool}$, are shown in the bottom panel of Figure \ref{fig:1d_torus_plots}.
$\phi_{\rm BH}$ saturates at around $\sim 40$ by $40{, }000M$ and its variability is almost entirely driven by variability in $\dot M $.  
That is, the un-normalized magnetic flux $\Phi_{\rm BH}$ is roughly constant with time for a majority of the simulation.  
We find that the integrated cooling rate is highly correlated with the accretion rate and thus displays similar variability, as expected in a \citet{Shakura1973} disk.

We quantify how well we are resolving the MRI in our simulations by using the $Q_{(i)} = \lambda_{{\rm MRI},i}/\Delta x_{(i)}$ parameter, where (e.g., \citealt{Noble2010})
\begin{equation}
\label{eq:mri_lambda}
\lambda_{{\rm MRI},i} = \frac{2 {\rm \pi} }{\sqrt{\rho} \Omega} b_\mu \hat e^{\mu}_{(i)}
\end{equation}
is the fastest growing MRI wavelength in the $i \in \{r,\theta,\phi\}$ direction, $\Delta x_{(i)} = dx^\mu e_\mu^{(i)}$, $\Delta x^\mu = (\Delta t, \Delta x, \Delta y, \Delta z)$ is a four vector constructed from the local cell size in the simulation frame, and $e^\mu_{(i)}$ is a set of orthonormal tetrads representing the zero angular momentum (ZAMO) frame (e.g., \citealt{McKinney2012}) converted from spherical Kerr-Schild coordinates to Cartesian Kerr-Schild coordinates.  
We plot the magnetic pressure-weighted averages of $Q_\theta$, $Q_\varphi$, and $Q_{\rm r}$ in the middle panel of Figure \ref{fig:1d_torus_plots} for $t=70{, }000M$ and $t=83{, }000M$, where we have restricted the averages to be within one scale height of the midplane.
Until $t=70{, }000M$, the MRI is well resolved, with $Q_\theta \gtrsim 15$ and $Q_\varphi\gtrsim 60$--100.
After this point, however, the vertical and radial component of the field decay in the main body of the disk to the point that $Q_\theta$ dips to $\lesssim 6$ over $\sim$ 18--40$\rg$ and $Q_r$ does the same over $\sim$ 30--40$\rg$.
This drop in MRI quality factors is also associated with a drop in accretion rate as seen in the bottom panel of Figure \ref{fig:1d_torus_plots}, as well as the growth of non-axisymmetric over-density features.\footnote{These over-densities are likely a result of the \citet{Rossby1939} wave instability (RWI, see \citealt{Lovelace2014} for a review in astrophysical contexts), which is triggered when a disk develops a sharp density gradient as occurs in, e.g., galactic disks \citep{Lovelace1978}, protoplanetary disks \citep{Varniere2006}, circumbinary disks \citep{Mignon2023}, and collapsars \citep{Gottlieb2024}. 
Here, magnetic truncation causes the mass density to peak at $10$--15$\rg$, allowing the instability to grow.  
The decline of $Q_\theta$ after $
\sim 70{ ,}000M$ in our simulations may then be a physical feature rather than a limitation of numerical resolution.
Regardless, we reserve an analysis of the long-term evolution of the RWI in truncated disks to future work. 
}
We thus use $t=50{, 000}M$ as the starting point for our binary simulations in order to have $\sim 20{, }000M$ of runtime while the MRI is still resolved.

\begin{table}
  \begin{center}
    \caption{Location of refinement levels in our single black hole disk simulations.  
    In addition to the zones listed here, there are also refinement levels added by the simulation to ensure that any given meshblock is never bordered by another meshblock with more than one level above or below its refinement level.      }
    \label{tab:resolution}
    \def\arraystretch{1.75}
    \begin{tabular}{|c|c|c|c|} 
            \hline
             Level & $x,y$-range & $z$-range  &  $\Delta z = \Delta x/3= \Delta y /3$\\
            \hline
                  $0$ & $ |x,y| \lesssim 384 \rg$ & $|z| \lesssim 384\rg$ & 2.7 $\rg$\\
      \hline
       $1$ & $ |x,y| \lesssim 256 \rg$ & $|z| \lesssim 98\rg$  & 1.3 $\rg$ \\
       \hline
       $2$ & $ |x,y| \lesssim 128 \rg$ & $|z| \lesssim 49\rg$  & 0.67 $\rg$ \\
       \hline
        $3$ & $ |x,y| \lesssim 96 \rg$ & $|z| \lesssim 24.5\rg$  &0.33 $\rg$ \\
               \hline
        $4$ & $ |x,y| \lesssim 80 \rg$ & $|z| \lesssim 12.25\rg$  &0.17 $\rg$ \\
        \hline
        $5$ & $ |x,y| \lesssim 47.5 \rg$ & $|z| \lesssim 6.125\rg$  &0.083 $\rg$ \\
        \hline
        $6$ & $ |x,y| \lesssim 23.75 \rg$ & $|z| \lesssim 3.0625\rg$  &0.042 $\rg$ \\
        \hline
        $7$ & $ |x,y| \lesssim 12 \rg$ & $|z| \lesssim 1.53125\rg$  &0.021 $\rg$ \\
        \hline
    \end{tabular}
  \end{center}
\end{table}

\section{Adding the Secondary Black Hole}
\label{app:metric_change}

The addition of the secondary black hole to our simulations is performed by instantaneously changing the spacetime metric.
To make the total energy $E=t^\mu_t$ consistent with this change, we add a one-time source term equal to 
\begin{equation}
    \Delta E = \frac{1}{2}\left[g_{\mu \nu}^{(new)}-g_{\mu\nu}^{(old)}\right]T^{\mu \nu}.
\end{equation}
Moreover, to account for the instantaneous change in volume and area in each cell, we then scale all the conserved quantities and magnetic fields by a factor of $\sqrt{-g_{\rm old}}/\sqrt{-g_{\rm new}}$ \citep{Ressler2024}.

\section{Dependence On The Target Scale Height}
\label{app:hr_03}
\begin{figure}
\includegraphics[width=0.47\textwidth]{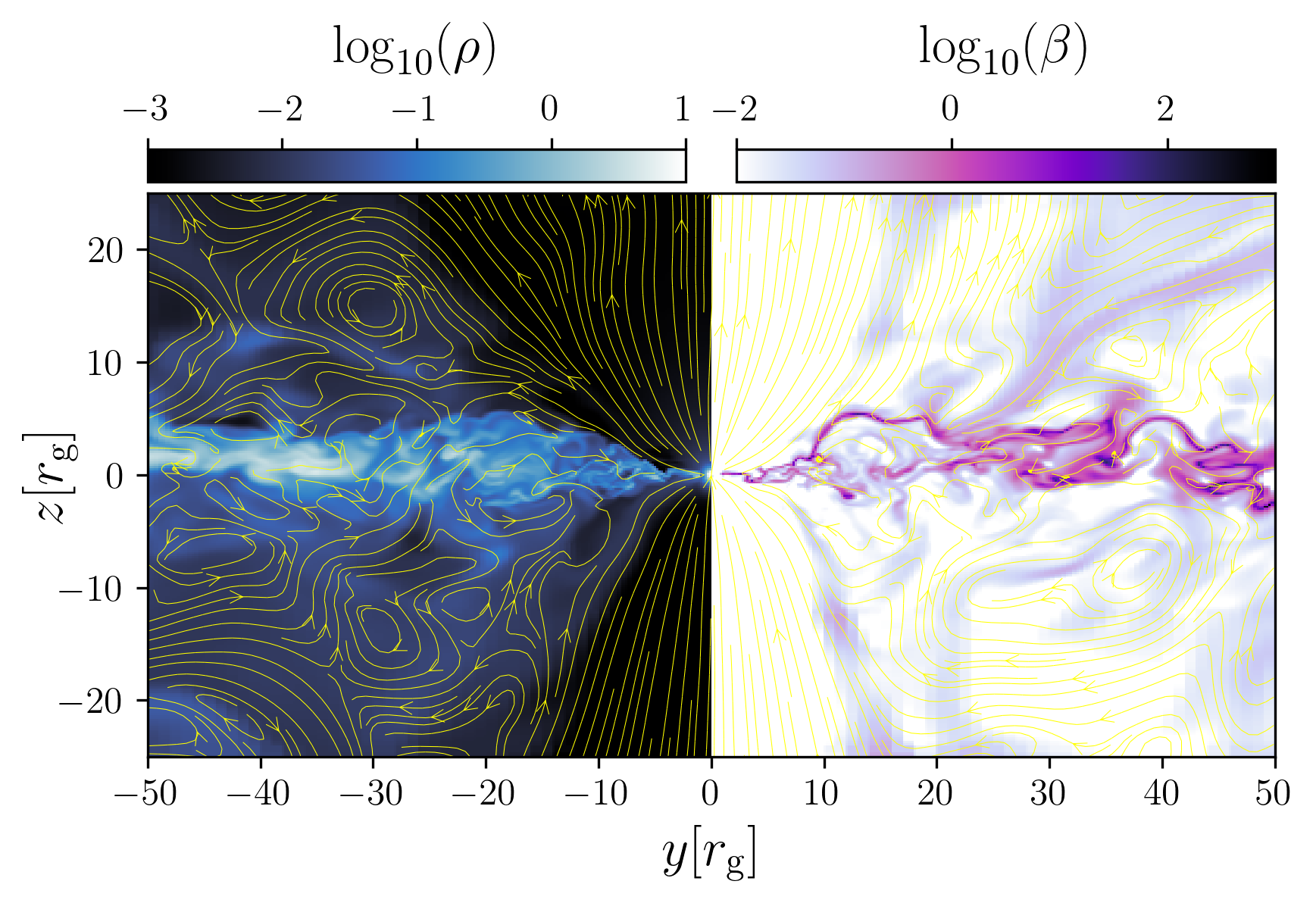}
\caption{Poloidal slice of density, $\rho$, (left) and plasma $\beta$ (right) with magnetic field lines overplotted at $t=70{, }000M$ from our single black hole, thin disk simulation that was restarted from the $H/r=0.1$ simulation to have a new $H/r=0.03$ target.
The much thinner disk is now highly magnetized due to compression of the magnetic field, with a lower density and faster inflow velocity.}
\label{fig:final_state_h_r_03}
\end{figure}

In order to test the dependence of our results on the target $H/r$, we restart our $H/r=0.1$ single black hole simulation at $t=50{, }000 M$ with a new $H/r=0.03$ target and run for an additional $20{, }000M$.  
By this time the disk has had time to cool significantly and compress, as shown in Figure \ref{fig:final_state_h_r_03}, where we plot $y$-$z$ contours of mass density and plasma $\beta$ overplotted with magnetic field lines.  
The disk is now highly magnetized, with $\beta \lesssim 1 $, and the structure is much more turbulent than the $H/r=0.1$ disk.  

\begin{figure}
\includegraphics[width=0.47\textwidth]{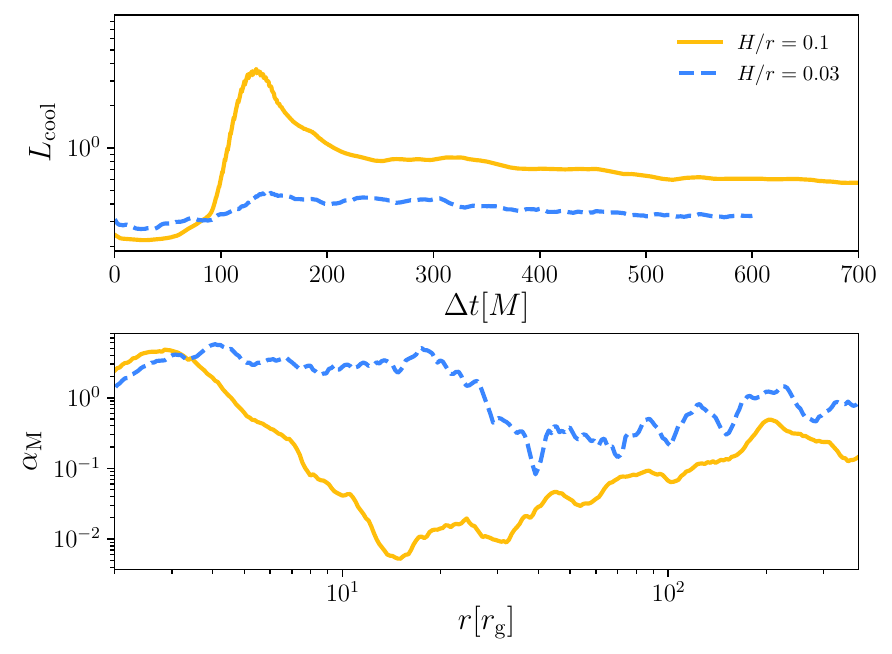}
\caption{
Comparison between our fiducial $q=0.1$, $H/r=0.1$ simulation (solid gold line) and our restarted, $q=0.1$, $H/r=0.03$ simulation (blue dashed line).
Top: Volume-integrated cooling rate, $L_{\rm cool}$, vs.\ time.
Bottom: Co-moving frame Maxwell stress, $\alpha_{\rm M}$, vs.\ radius just before the secondary is introduced.
Despite the thinner primary disk, the impacts from the secondary in the $H/r=0.03$ simulation are relatively much less powerful than in the $H/r=0.1$ simulation, where there is a clear peak in $L_{\rm cool}$ that stands out from quiescence. 
This is because of the much higher (by a factor of $\gtrsim 100$) $\alpha_{\rm M}$ in the $H/r=0.03$ simulation caused by the rapid compression of the magnetic field and reduction in pressure.
This makes the quiescent disk heating relatively stronger than in the $H/r=0.1$ simulation.
While this highlights the importance of the accretion state in modeling OJ 287-like systems, we caution in the main text that this state may be artificial.
}
\label{fig:hr_comp}
\end{figure}

To this simulation, we then introduce a secondary black hole with $q=0.1$ in the same way as the simulations described in the main text.  
We run this simulation only for a short time to capture the first impact of the secondary with the disk.  
The resulting integrated cooling rate is plotted vs.\ time in the top panel of Figure \ref{fig:hr_comp} compared to the $q=0.1$, $H/r=0.1$ simulation.
Surprisingly, the impact in the $H/r=0.03$ simulation results in a barely noticeable peak in $L_{\rm cool}$ relative to quiescence compared with a very clear peak for $H/r=0.1$.
This seems counterintuitive because the analytics (Equation \ref{eq:heating_ratio}) predict that the ratio between the total impact heating and quiescent disk heating should scale as $q^2/(H/r)^3$.  

The apparent discrepancy is resolved by considering the fluid frame Maxwell stress in the two simulations, $ \alpha_{\rm M} = -\langle b^r b_\varphi \rangle/\langle P\rangle$, where $b^r$ and $b_\varphi$ are the radial and azimuthal components of the covariant and contravariant four magnetic field in spherical Kerr-Schild coordinates converted from Cartesian Kerr-Schild coordinates. 
This quantity is plotted in the bottom panel of Figure \ref{fig:hr_comp}, where the averages are restricted to within one scale height of the midplane in each case.
Because of both the enhanced magnetic field strength from compression and the reduced pressure support caused by cooling, $\alpha_{\rm M}$ is now $>100$ times larger in the $H/r=0.03$ simulation than in the $H/r=0.1$ simulation at the radius of first impact, $\sim 20 \rg$.
This reduces the relative importance of the total impact heating rate by a larger factor than it is increased by the reduction in scale height, as $Q_{\rm s}/Q_{\rm disk}$ is expected to scale as $1/\alpha$ (Equation \ref{eq:heating_ratio}).
In fact, because of this substantially increased Maxwell stress in the $H/r=0.03$ simulation, the accretion rate is roughly the same as in the $H/r=0.1$ simulations, with a lower density midplane and higher inflow velocity.
As a result, the impacts from the secondary black hole have a much weaker effect on altering the disk dynamics and have less relative power to outshine the disk.

There are several reasons, however, to doubt whether our $H/r=0.03$ simulation is realistic.
First, the simulation is only run for $20{, }000M$ after the $H/r$ target is reduced.  
Our $H/r=0.1$ disk also displayed a transient, highly magnetized $\beta \lesssim 1$ phase during the first $\sim 20$--$30{ ,}000M$ before settling down into a much more weakly magnetized $\beta \gtrsim 100 $ disk. 
Since the dynamical timescales in an $H/r=0.03$ disk are naively expected to be much longer, we anticipate that the same will eventually happen in that simulation if we were to evolve it for a longer time.
Second, because we do not also increase the resolution when reducing the target $H/r$, the new disk is less resolved by a factor of $\sim$3, with 10--20 cells per scale height. 
This is below the recommended $\sim$ 30 cells per scale height to properly resolve the disk dynamics \citep{Noble2012}.
Third, it is unclear whether simply cooling a thicker magnetized disk is a realistic initial condition for these systems.
This process likely artificially enhances the relative magnetic field strength through rapid compression, biasing the evolution towards lower $\beta$.   
Determining more realistic initial conditions for AGN is still an active area of research (e.g., \citealt{Ressler2020b,Ressler2023,Cho2023,Guo2024,Hopkins2024,Kaaz2025,Guo2025,Wang2025}) .

Despite these uncertainties, the results presented in this Appendix highlight the importance of the accretion state of the primary on constraining observables.
For systems like OJ 287, it is thus critical to know how AGN disks in the relevant Eddington ratio range behave using radiative GRMHD simulations with realistic initial conditions.

\section{Sensitivity to The Cooling Function}
\label{app:cooling_function}

The cooling function we use in our simulations, $\Lambda_{\rm cool}$, is artificial in that it is not based on any physical radiation mechanism.  
Instead it simply drives the gas towards a targeted temperature associated with a constant $H/r$ in the accretion disk (Equation \ref{eq:lambda_cool}).
While the particular functional form has become widely used since its first appearance in the literature \citep{Noble2009}, it was chosen somewhat arbitrarily to ensure numerical stability and rapid cooling.
On the surface, this would seem to severely limit the applicability of any volume-integrated cooling rate to observable systems.
However, because our simulations are energy conserving, if the cooling occurs more rapidly (defined with respect to the cooling time $u_{\rm g}/\Lambda_{\rm cool}$) than the heating, the temperature (and to a lesser extent, other variables like density) will adjust so that $L_{\rm cool} = Q_{\rm heat}$ on timescales comparable to the cooling time, where $L_{\rm cool}$ and $Q_{\rm heat}$ are the volume-integrated heating and cooling rates, respectively.  
This is certainly true in typical quasi-steady thin disk simulations like the one we use as initial conditions described in Appendix \ref{app:initial_condition}.
In that case, the heating time can be orders of magnitude longer than the orbital time (by a factor of $\sim 1/\alpha $ where $\alpha$ is the effective viscosity parameter) while the cooling time is some fraction of the orbital time.  
As a result, $L_{\rm cool}$ is then a reliable estimate of the integrated heating rate.
This can be seen in the bottom panel of Figure \ref{fig:1d_torus_plots}, where $L_{\rm cool}$ is tightly correlated with the accretion rate, agreeing with the $\alpha$-disk theory expectation of $L \propto \dot M$.

For transient heating events like the impact of the secondary black hole with the primary disk in the simulations presented in the main text, we must verify whether or not the cooling time is short enough for $L_{\rm cool}$ to faithfully represent the intrinsic heating rate.
Analytically, we expect this heating time to be $t_{\rm heat} \approx H/v_{\rm orbit} \approx (H/r)/\Omega_{\rm orbit}$, where $v_{\rm orbit}$ and $\Omega_{\rm orbit}$ are the linear and angular velocity of the orbit, respectively.  
The cooling time in Equation \ref{eq:lambda_cool} is $\sim (T_{\rm  target}/T)^{1/2}/\Omega_{\rm K}$ for $T\gg T_{\rm target}$. 
For a strong shock, and for the gas at distances greater than $q r_{\rm orbit}$ from the secondary, the post-shock temperature $T$ can be related to the velocity of the secondary to obtain $t_{\rm cool} \sim (H/r)/\Omega_{\rm orbit}$, which is $\approx t_{\rm heat}$. 
For distances within $q r_{\rm orbit}$ from the secondary, the cooling time is much shorter in Equation \eqref{eq:lambda_cool}, $t_{\rm cool} \sim (H/r)/(q\Omega_{\rm orbit})$.
This means that our integrated $L_{\rm cool}$ could potentially fail to account for some of the heating caused by the secondary-induced shock if a significant amount of heating occurs at distances greater than $q r_{\rm orbit}$ from the secondary.

\begin{figure}
\includegraphics[width=0.47\textwidth]{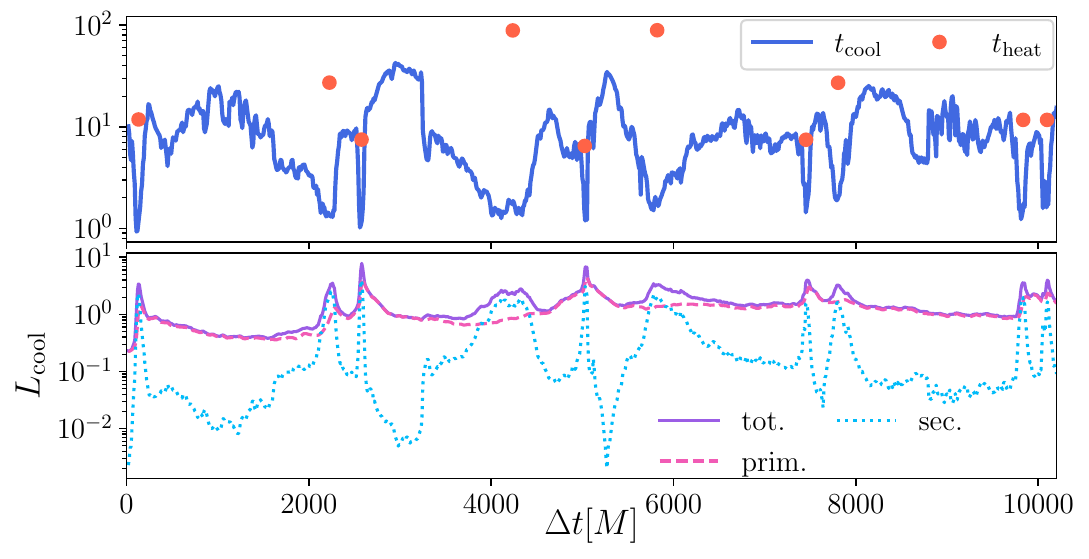}
\caption{Top: Heating time (or disk crossing time), $t_{\rm heat} = H/v_{\rm orbit}$ (red dots), for each secondary impact and cooling function-weighted volume average of the cooling time, $t_{\rm cool}$, vs.\ time (solid blue line).
Bottom: Volume-integrated cooling rate, $L_{\rm cool}$, vs.\ time (solid purple), broken up into the contribution by the secondary cooling function (within $q r_{\rm orbit}$ of the secondary, dotted blue line) and primary cooling function (everywhere else, dashed pink line).
During impact events, the cooling function is typically dominated by the secondary contribution and the associated cooling time is always small compared with the heating/disk crossing time.  
This means that the integrated cooling rate can accurately measure the intrinsic heating rate induced by the secondary impacts.
}
\label{fig:heating_time}
\end{figure}

To determine whether or not this is the case, we can compute the average cooling time directly from our simulation, weighted by the cooling function:
\begin{equation}
\label{eq:tcool}
    \langle t_{\rm cool}\rangle = \frac{1}{L_{\rm cool}}\int\int \int  - \left(\frac{u_{\rm g}}{\Lambda_{\rm cool}}\right)\Lambda_{\rm cool} u_t \gdet dx dy dz.
\end{equation}
This quantity is plotted in the top panel of Figure \ref{fig:heating_time} along with the disk crossing times/shock heating times for each secondary impact (more precisely, when the secondary crosses the midplane) for our $q=0.1$ simulation.
We find that the cooling time is always less than the heating/crossing time during the impacts, typically by a factor of 10 or more.  
The reason for this is because the cooling during these impacts is often dominated by the secondary cooling function (that is, the cooling function that turns on within $qr_{\rm orbit}$ of the secondary that has a shorter cooling time by a factor of $q$). 
This can be seen in the bottom panel of Figure \ref{fig:heating_time}, where we plot $L_{\rm cool}$ for the same simulation broken down into the `primary' and `secondary' components.  
During flares/impacts, the contribution to the total cooling rate is briefly dominated by the secondary cooling function while at all other times the total cooling function is dominated by the primary cooling function.
This means that our measurement of $L_{\rm cool}$ in our simulations is likely an accurate representation of the intrinsic heating rate.

\begin{figure}
\includegraphics[width=0.47\textwidth]{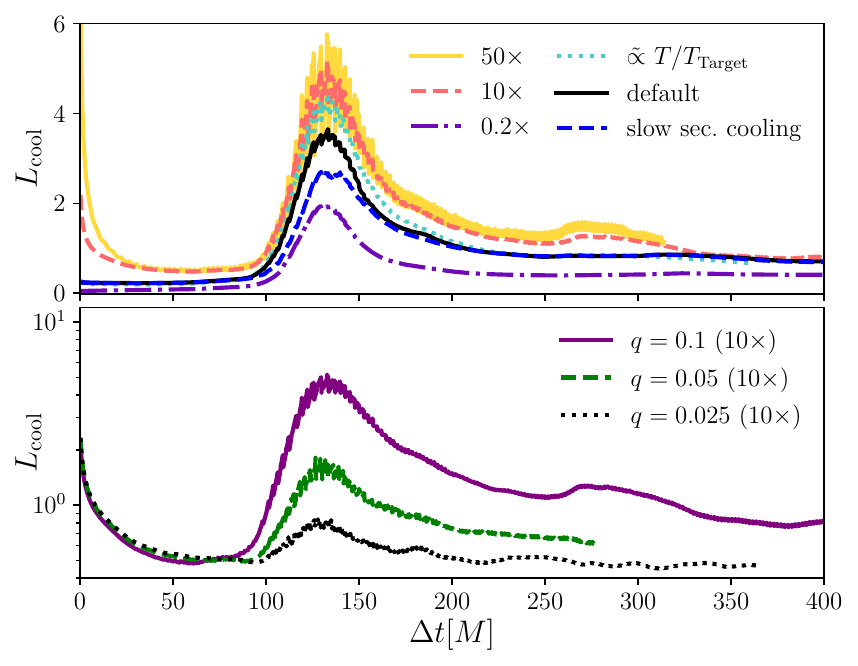}
\caption{Top: Volume-integrated cooling rate vs.\ time during the first impact of the secondary black hole with the primary accretion disk for different forms of the cooling function.
Bottom: Integrated cooling rate for a cooling function $10$ times larger than our default cooling function for the three value of $q$ used in the main text. 
These plots show that the default cooling function produces an $L_{\rm cool}$ that is close to the saturated value representing the 'true' heating rate induced by the secondary impact, and that the dependence of $L_{\rm cool}$ on $q$ is not sensitive to the cooling function (compare with Figure \ref{fig:Lcool_vt}).  
}
\label{fig:cooling_comp}
\end{figure}

To further establish this claim, we also ran several additional simulations through the first impact of the secondary using different forms of the cooling function.
In particular, we ran simulations using cooling functions multiplied by constant factors of 0.2, 10, and 50 everywhere, as well as a cooling function that depends on $(Y-1 + |Y-1|)$ instead of $(Y-1 + |Y-1|)^{1/2}$ (see Equation \ref{eq:lambda_cool}), and a cooling function that does not include the factor of $q$ in the secondary portion of the cooling function (i.e., the cooling is slower when close to the secondary compared to our default cooling function).  
The integrated cooling rates for all of these are plotted in the top panel of Figure \ref{fig:cooling_comp}.
Generally, we find that the cooling functions with faster cooling times essentially converge to the same integrated value, which is $\sim 30\%$ higher than that obtained with our default cooling function at the peak and likely represents the `true' intrinsic heating rate.  
On the other hand, shortening the cooling time can make $L_{\rm cool}$ arbitrarily small, so that it is significantly more sensitive to constant factors that reduce the cooling function than constant factors that increase it (when close to the saturated cooling rate that represents the ``true'' heating rate).  
Similarly, the reduction of the secondary contribution to the cooling function compared to the default cooling function causes $L_{\rm cool}$ to further underestimate the intrinsic heating rate.
In conclusion, we find that $L_{\rm cool}$ calculated using our default cooling function in Equation \ref{eq:lambda_cool} is only modestly underestimating the intrinsic heating rate in the simulation and captures its qualitative behavior well. 

Finally, to check whether or not the dependence of the measured $L_{\rm cool}$ in our simulation on $q$ is sensitive to the cooling function, we also plot $L_{\rm cool}$ from simulations that use the $10$ times higher cooling function for $q=0.1$, $q=0.05$, and $q=0.025$ in the bottom panel of Figure \ref{fig:cooling_comp}.  
We find the same rough scaling with $q$ using the more rapid cooling function as we do in the default cooling function (see Figure \ref{fig:Lcool_vt}). 
Namely, $L_{\rm cool}$ has a steeper than $q$ but shallower than $q^2$ dependence.
This provides reasonable evidence that the dependence of $L_{\rm cool}$ on $q$ measured in our simulations is robust.

\section{Gravitational Torques On The Primary Disk}
\label{app:torques}

Our simulations demonstrate that gravitational torques from the secondary black hole can significantly tilt the primary accretion disk.  
To understand this more clearly, we consider a ring of gas with uniform density at some distance from the primary in the Newtonian limit.  
The ring is two-dimensional and occupies the $x$-$y$ plane with angular momentum in the $+z$ direction.  
We then consider a secondary orbiting clockwise in the $y$-$z$ plane with orbital angular momentum in the $-x$ direction.  
The torque exerted on an individual parcel of gas in the ring is $\mathbf{r} \times F_{\rm g}$, where $F_{\rm g} = -dm (\mathbf{r} - \mathbf{r_{\rm s}})/|\mathbf{r} - \mathbf{r_{\rm s}}|^3$, $dm$ is the mass of a gas parcel, and $\mathbf{r_{\rm s}}$ is the location of the secondary.  
For any given parcel of gas and at any given time, because the orbit is in the $y$-$z$ plane the torque in the $y$ direction always sums to zero over the whole ring as long as the ring remains in the $x$-$y$ plane.
However, there can be significant torque in the $\pm$$x$-direction that does not average to zero over the disk at a fixed time.
Although this torque approximately averages to zero over a single orbit of the secondary, on shorter timescales it can tilt the disk in either the $+x$ or $-x$ direction, at which point the $y$ component of the torque no longer averages to zero over the disk because the symmetry has been broken.  
The subsequent evolution of the angular momentum and ring direction is nonlinear and thus somewhat unpredictable, especially in a more realistic system (though see \citealt{Ivanov2024} for an analysis for $H/r \sim 10^{-3}$ and $\alpha=0.1$ viscosity).
This is because during a single orbit the secondary causes both positive and negative torques, so the disk doesn't monotonically tilt in any direction.
Furthermore, in addition to any asymmetries that exist initially in the disk structure, the impacts of the secondary induce highly asymmetric density features that complicate the simple picture of a uniform disk. 
Long-term simulations will be needed to understand the ultimate fate of an accretion disk being impacted by an inclined secondary black hole.

\end{document}